\documentclass{article}

\usepackage{arxiv}
\usepackage{amssymb}
\usepackage{amsmath}
\usepackage[noend]{algorithm2e}
\usepackage[utf8]{inputenc} 
\usepackage[T1]{fontenc}    
\usepackage{hyperref}       
\usepackage{url}            
\usepackage{booktabs}       
\usepackage{amsfonts}       
\usepackage{nicefrac}       
\usepackage{microtype}      
\usepackage{parskip}
\usepackage{lipsum}         
\usepackage{graphicx}
\usepackage{natbib}
\usepackage{doi}
\usepackage{hyperref}
\usepackage{float}
\usepackage{verbatim}
\usepackage{tabularx}
\usepackage{multirow}
\usepackage{setspace}
\usepackage{subcaption}

\DeclareMathOperator*{\argmax}{argmax}

\newcolumntype{Y}{>{\centering\arraybackslash}X}
\newcommand{\tabref}[1]{Table \ref{#1}}
\newcommand{\equref}[1]{Eq. \ref{#1}}
\newcommand{\figref}[1]{Fig. \ref{#1}}
\newcommand{\secref}[1]{Section \ref{#1}}
\newcommand{\algref}[1]{Algorithm \ref{#1}}

\title{DANI: Fast Diffusion Aware Network Inference with Preserving Topological Structure Property}

\author{%
    Maryam Ramezani \\
    Department of Computer Engineering \\
    Sharif University of Technology \\
    \texttt{maryam.ramezani@sharif.edu} \\
    \And
    Aryan Ahadinia \\
    Department of Computer Engineering \\
    Sharif University of Technology \\
    \texttt{aryan.ahadinia@sharif.edu} \\
    \And
    Erfan Farhadi \\
    Department of Computer Engineering \\
    Sharif University of Technology \\
    \texttt{erfan.farhadi76@sharif.edu} \\
    \And
    Hamid R. Rabiee \\
    Department of Computer Engineering \\
    Sharif University of Technology \\
    \texttt{rabiee@sharif.edu} \\
}

\date{}

\renewcommand{\shorttitle}{DANI}

\hypersetup{
pdftitle={DANI: Fast Diffusion Aware Network Inference with Preserving Topological Structure Property},
pdfsubject={cs.SI},
pdfauthor={Maryam Ramezani, Aryan Ahadinia, Erfan Farhadi, Hamid R. Rabiee},
pdfkeywords={First keyword, Second keyword, More},
}

\begin{document}
\maketitle

\begin{abstract}
    The fast growth of social networks and their data access limitations in recent years has led to increasing difficulty in obtaining the complete topology of these networks. However, diffusion information over these networks is available, and many algorithms have been proposed to infer the underlying networks using this information. The previously proposed algorithms only focus on inferring more links and ignore preserving the critical topological characteristics of the underlying social networks. In this paper, we propose a novel method called DANI to infer the underlying network while preserving its structural properties. It is based on the Markov transition matrix derived from time series cascades, as well as the node-node similarity that can be observed in the cascade behavior from a structural point of view. In addition, the presented method has linear time complexity (increases linearly with the number of nodes, number of cascades, and square of the average length of cascades), and its distributed version in the MapReduce framework is also scalable. We applied the proposed approach to both real and synthetic networks. The experimental results showed that DANI has higher accuracy and lower run time while maintaining structural properties, including modular structure, degree distribution, connected components, density, and clustering coefficients, than well-known network inference methods.
\end{abstract}

\keywords{Network Science \and Social Networks \and Network Inference \and Diffusion Information \and Topological Structure }


\section{Introduction}\label{sec:introduction}

\begin{figure}[ht]
	\centering
	\includegraphics[width=0.4\textwidth]{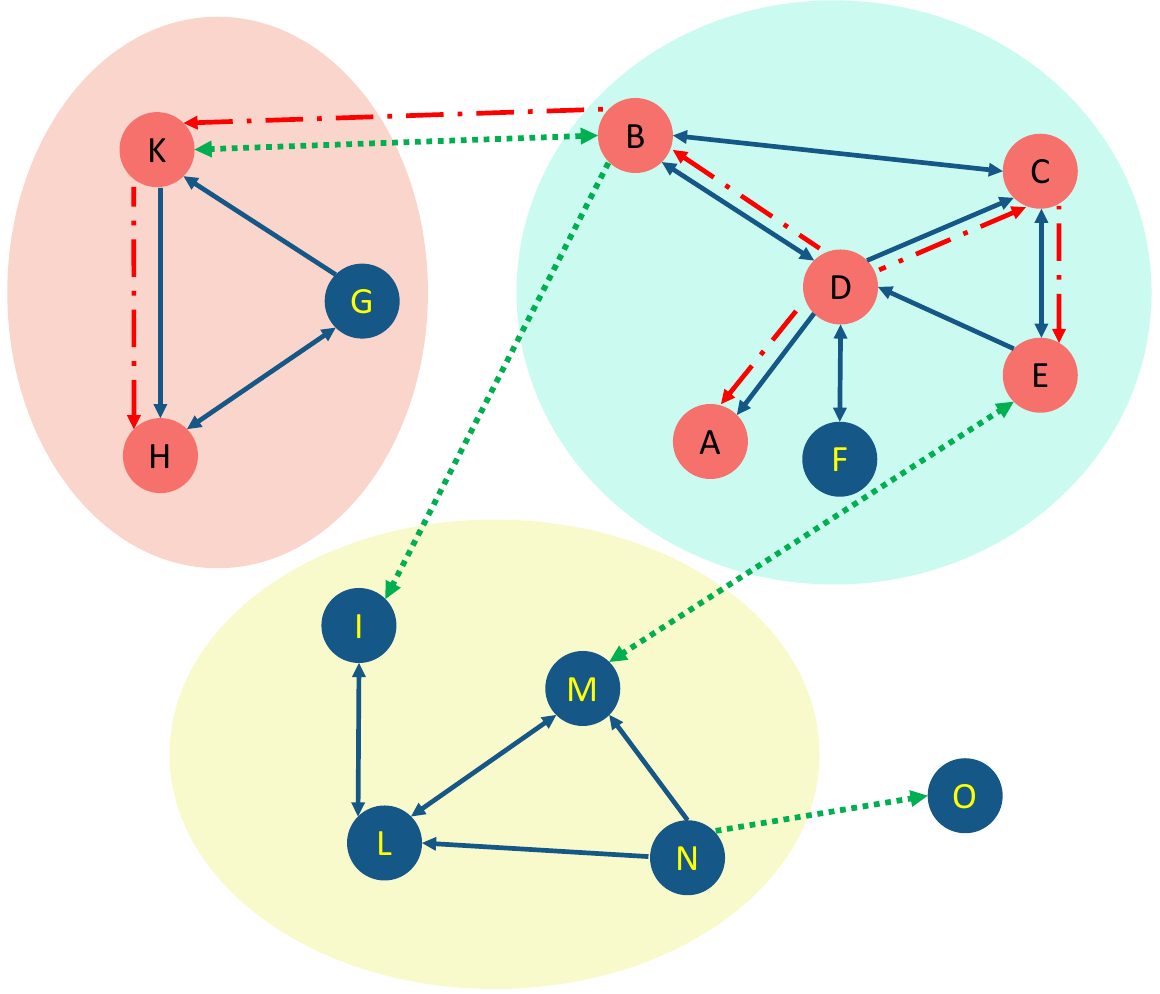}
	\caption{A toy example of one diffusion process over structural properties of the network (links and communities). Colored shadows separate the three communities. Strong ties (navy blue solid lines) form a community, and weak ties (green dash lines) act as bridges between different communities. Core nodes ( $\{A,C,D,F,G,H,L\}$), and boundary nodes ($\{B,E,I,K,M,N,O\}$) are shown. The directed path (red directed lines and nodes) shows a typical cascade over the network.}
	\label{fig:cascade}
\end{figure}

In today's world, Online Social Networks (OSNs) play a significant role in the exchange of information. Information, ideas, and behaviors are disseminated over these networks through \textit{diffusion}, namely \textit{contagion}. 
As shown in \figref{fig:cascade}, the propagation of contagion over a network creates a trace that is called \textit{cascade} \cite{NetInf:2010}. 
Diffusion cascades provide valuable information about the underlying networks. 
Based on previous studies, diffusion behavior, and network structure appear tightly related \cite{kumar2021information,easley2010,Eftekhar:2013}. 
Therefore, users' actions are driven by their interests and influenced by interaction patterns. A better understanding of this relationship will allow a more accurate analysis of social networks and their processes (e.g., diffusion). 
In many situations, the underlying network topology (nodes and the links between them) may not be achievable for different reasons:

\begin{enumerate}
    \item Privacy and protected accounts
    \item Information published on the websites without referencing each other 
    \item Limited access to data due to the policies of OSNs
\end{enumerate}    

Network topology inference is a solution to surmount these difficulties.
    
As mentioned before, the diffusion process is influenced by the structure of the latent underlying network. 
Thus, observing nodes' diffusion behavior can help us gain insight into the network's structure. 
In most cases, the only available observation is the time when a contagion reaches a node, namely the infection time of a node. 
Network inference includes a class of studies that aim to discover the latent network by utilizing these infection times in different contagions.

According to previous research, social networks have unique features that differentiate them from random graphs. 
These characteristics among nodes and links include degree distribution, connected components, density, and clustering coefficients \cite{Interestclustering2021,kwak2010twitter}. The inferred and underlying networks should have similar nodes degree distributions, so influence nodes remain at high degrees and the inferred network can be used for tasks such as influence maximization for advertising goals.
One of the most important structural features of OSNs is their \textit{modular structure}. 
Each module in the network is a dense sub-graph called a \textit{community}. 
These properties have an impact on network interactions as well as how information spreads over them. 
A reliable inference must preserve the structural features to correctly reconstruct the network.

\textbf{Motivation}: Many structural services are provided on OSNs. 
Since gathering information about the entire network topology is often impossible, inferred networks from observations are usually used.
By preserving the topological structure during network inference, structural services such as optimizing search engines, improving recommendation systems, compressing network data, and viral marketing can be applied to the inferred network, leading to comparable results with the underlying network. In light of these concerns, we pose the following question:

\begin{quote}
\textit{How to accurately and efficiently infer a network from observed diffusion so that the properties of the inferred network will be similar to the underlying network properties?}
\end{quote}

Several network inference algorithms have been proposed, but most of them do not preserve network structural topology during inference \cite{PBI2023}. 
Despite their acceptable accuracy in detecting network links, the inferred network's structural properties do not match the underlying network. 
In this paper, we propose a new Diffusion Aware Network Inference algorithm, called \textit{DANI}, which infers social networks from diffusion information while persevering the topological structure. 
Based on the infection times of the nodes in different contagions, we estimate the network structure and evaluate how many structural properties are preserved by the inference.

The main contributions of this paper are:

\begin{itemize}
	\item \textbf{Inferring networks while preserving topological properties}: Different structural-based methods may use the inferred networks as an alternative to obtaining the actual network structure. 
	The previous network inferring works do not address the need to have the same link density, degree distribution, and clustering coefficient as the underlying network. Consequently, using community detection methods on their output deviates from community detection on the underlying network. 
	The proposed method overcomes this problem by preserving structure while inferring networks.
	\item \textbf{Inferring networks with more connected nodes}: Network nodes in inference problems are the active nodes participating in at least one observed cascade. Hence, any active node must have at least one connected link to enable cascade propagation through that link. 
	In contrast with previous research, the proposed method produces inferred networks with more connected nodes, similar to the underlying network.
	In other words, our method produces fewer isolated nodes than the previous methods.
	\item \textbf{More scalability}: 
	The proposed method is a parallelizable algorithm that can be used for big data. 
	It offers a lower run time than the state-of-the-art methods while maintaining high precision. 
	Hence our method is scalable, which is the main limitation of the previous works.
	\item \textbf{Good performance independent of the network structure}: 
	Many of the previous works have a good performance on tree-based networks, while the performance of the proposed method is not limited to any type of structure.
\end{itemize}

The rest of this paper is organized as follows. 
In \secref{sec:preliminaries}, we introduce the preliminaries on the concepts of structural properties and diffusion. Moreover, we explain their interdependencies to ascertain our concerns about preserving network topology during the inference process. Related works are presented in \secref{sec:related}. 
In \secref{sec:formulation}, we define notations and formalize our problem then in \secref{alg:DANI} we explain our method, DANI, and theoretically analyze scalability of the proposed method. 
Experimental results on both synthetic and real datasets are presented in \secref{sec:experiments}. Finally, we conclude the paper and discuss future works in \secref{sec:conclusion}.

\section{Preliminaries on diffusion and modular structure}\label{sec:preliminaries}

The role that different nodes and links play in network processes, such as diffusion, varies and can be described as \cite{journals/abs-1110-5813}: 

\begin{itemize}
	\item \textit{Intra-community links} that connect nodes in the same community. These nodes interact frequently; therefore, these links are known as strong ties. The nodes that only have links to other nodes in their community are called core nodes. The core nodes tend to influence their cohorts in a community more than other network members.
	
	\item \textit{Inter-community links} which connect nodes of different communities. Due to low interactions between such nodes, they are called weak ties. Nodes that have at least one inter-community link are called boundary nodes. Such nodes play an important role in information dissemination over different communities.
\end{itemize}

Analyzing the interactions between network members is an ongoing issue in social networks \cite{FuzzySystemsChen}. In fact, the probability that a person will reject or accept a contagion depends on the decision of his neighbors about that contagion. At the same time, this probability affects the future decisions of neighborhood nodes in the same community \cite{scripps2007roles}. Community members are exposed to contagion through strong intra-community links within a short period. In contrast, due to weak inter-community links between communities, contagion propagation occurs with a lower probability and pace. 

Another interesting finding is that in many cases, the impact of weak ties on a diffusion process (i.e., the acceptance rate of a new product) is more powerful than strong ties \cite{abs-1201-4145,goldenberg2001talk}. We try to explain the above facts through an example. A network with three communities is illustrated in \figref{fig:cascade}. Suppose that node $D$ gets infected by a contagion. Other nodes in the community of node $D$ will be exposed to this contagion and will become infected with high probability due to similar interests and actions among them. Thus, the rate and extent of diffusion can be high within a community. But how does the contagion disseminate over the network? For this purpose, the contagion must spread out of the community by a boundary node such as $B$ and infect another boundary node like $K$ from a neighboring community, and this only happens if there exists at least one inter-community link such as $(B,I)$ or $(B,K)$. Therefore, weak ties have a critical role in the diffusion process over the network. The diffusion process may stop since interaction with weak ties is less probable. In other words, diffusion and community may act against each other. Communities may prevent a cascade, and in most cases, when a cascade stops, a community can be detected \cite{easley2010,BarbieriBM13}. 

Therefore, it is necessary to understand the relation between the community structure and diffusion behavior of nodes in a network to infer the network correctly.

\section{Related works}\label{sec:related}

As discussed in the previous sections, we would like to infer a network by utilizing diffusion information while considering the logical structure. 
Consequently, the most related research areas to our work are the methods that infer networks by observing the diffusion network, especially those considering the structural properties. In this section, we explain this research area.
The general goal is trying to infer the edges of a network by using cascade information, which in most cases is the infection time of nodes in different cascades.

There are comprehensive surveys such as \cite{li2021capturing,brugere2018network,zhou2021survey} about concepts of this area. In the following, we describe some of the most significant works related to the present article.\\

\textbf{Tree-based-Maximum-Likelihood-Estimation Methods}:
Special attention to this area began with NetInf, which models each cascade as a tree on the network and uses an iterative algorithm to infer the network from these trees by optimizing a sub-modular function \cite{NetInf:2010}. Despite its high inference accuracy, it suffers from high runtime, which is a serious problem in real datasets. In addition, the performance of NetInf decreases when the structure of the network being studied is not consistent with the tree model \cite{Fastinf11}. Similar to NetInf, MultiTree models cascade with the trees but consider all the possible trees \cite{journals/abs-1205-1671}.\\

\textbf{Convex-based-Maximum-Likelihood-Estimation Methods}:
CoNNIE improves NetInf by adding an optimal and robust approach that uses prior probabilistic knowledge about the relation between infection times \cite{nipsMyersL10}. NetRate is another improvement over NetInf. It assumes that cascades occur at different rates and temporally infers heterogeneous interactions with varying transmission rates, which is closer to reality \cite{Netrate11}. InfoPath \cite{conf/wsdm/Gomez-RodriguezLS13} is an extension of NetRate for dynamic networks. MultiTree's accuracy is higher than NetInf, NetRate, and CoNNIE when a few cascades are available because it considers all the possible trees. Although its running time is several orders lower than other tree-based algorithms, it is still not scalable. Some recent works try to consider structural properties. MADNI utilizes various triangle motif patterns as prior knowledge \cite{tan2020motif}. It infers the diffusion network by maximizing the likelihood of observed cascades with a regularization term on the mined motif structure. The inverse number of motif patterns containing a link is added as a penalty to regularize the model likelihood. However, MADNI is prone to mining closed triangle motifs and not considering other motifs patterns.

Unlike mentioned works which are based on infection time, TENDS only requires the infection status of nodes in each cascade and does not rely on the actual time or sequence of infection. TENDS tries to learn a model that succeeds in accurately inferring links while interested in low statistical error. It combines maximum likelihood and statistical error by using a scoring criterion as balance and solves it with a greedy search procedure using a Mutual Information based pruning method \cite{han2020statistical}.\\

\textbf{Markov-Random-Walk-based Methods}:
DNE models diffusion as a Markov random walk and tries to define a weight for each link using hitting time \cite{EslamiRS11}. 
FastINF is another method that uses the same approach to determine the weight of links \cite{Fastinf11}. 
These algorithms have a low runtime and work well on the networks but cannot preserve the topological structure. The main drawbacks of these methods are as follows:
(1) Algorithms like NetInf and others based on it, which use a spanning tree approach, maintain the structures but are too slow. These methods have low performance on networks with non-tree structures.
(2) Others like DNE and FastINF distinguish no preference between different types of extracted edges. Despite their high accuracy, they do not consider the network structure. Running community detection algorithms on the output of these methods leads to different communities from the underlying network because they do not recover some weak inter-community or strong intra-community links.\\

\textbf{Bayesian-Inference-based Methods}: 
\cite{gray2020bayesian} proposes a Bayesian inference method using Markov Chain Monte Carlo (MCMC) and the Tie No Tie (TNT) sampler for sampling from the posterior $P(G|C)$, with the likelihood of information propagation $P(C|G)$ over all directed spanning trees. In this method, the Bayesian part is based on an undirected graph. Due to its sampling method, its convergence is slow. PBI first runs DANI \cite{ramezani2017dani} to achieve a pairwise interaction network and converts it to an undirected network, then iteratively employs the TNT sampler and MCMC algorithm to add or remove a random edge to this network \cite{PBI2023}. Because of using DANI's output as the initial graph, it preserves modular structure and can detect edges in undirected graphs with high accuracy after Bayesian iterations. But if the graph is directed, the quality will be lost.\\

\textbf{Embedding-and-Clustering-based Methods}:
CENI clusters the nodes of cascades, and for nodes in the same cluster, it infers the links by utilizing survival functional same as NetRate \cite{hu2016clustering}. 
Assume that $i=(u,v)$ are pairs of two nodes that participate in at least $20$ cascades with the condition that node $u$ has a lower infection time than node $v$. 
A data group is a set of tuples $D^{(i)}=\{(x^{(i)}_{c},y^{(i)}_{c})\}$ where $y^{(i)}_{c}$ is the difference in infection time of pair nodes $i$ in cascade $c$, and $x^{(i)}_{c}$ is the description of related cascade $c$ that ${i}$ have participated in. KEBC embeds all possible data groups in a network using the Reproducing Kernel Hilbert Space method \cite{hu2019model}. By using K-means, it divides them into two clusters. It interprets this information to indicate links between nodes in a cluster with lower infection time differences, whereas nodes in the other cluster are not connected directly. Running KEBC takes a long time and depends on the parameters of embedding. In REFINE, the interaction pattern of nodes from multiple cascades is summarized into a sparse matrix, and the user embedding is constructed with TSVD and autoencoder \cite{kefato2019refine}. REFINE network reconstruction is based on cosine similarity calculated from embedded user's features. There are limitations to REFINE, such as its applicability to undirected graphs, reliance on hyperparameters which must be tuned through observation of the underlying network, and time and space complexity.

Here, we aim to infer the network by observing the cascade information with a fast and scalable method. In this regard, we desire to maintain the structural properties of the underlying network.

\section{Problem Formulation}\label{sec:formulation}

In this section, we first present the notations used in this paper and then introduce the proposed algorithm.
Let ${G^{*}}=(V,E)$ represent a network where $V$ is the set of nodes, and $E$ is the set of directed interaction links between nodes.

We assume to have a set of different contagions as different cascades,  $C=\{c_1,c_2,\ldots,c_M\}$. For each cascade $c_i$, we have pairs of node names and infection time, $c_i=\{(v_0,t^{(i)}_0),(v_1,t^{(i)}_1),\ldots,(v_n,t^{(i)}_n)\}$
where $t^{(i)}_0 < t^{(i)}_1 < \dots < t^{(i)}_n$. During the spread of a contagion, some nodes may not get infected. An infinity infection time is assigned to such nodes in that cascade. The propagation path of contagion is hidden. In other words, we do not know by which node a node gets exposed to contagion. It is assumed that a node can get infected with a certain contagion just by another node. 

In general, infection times form a continuous sample space, making their interpretation somewhat difficult. In addition, the range of infection time may vary for each contagion. For example, the order of infection time may be in seconds for cascade $c_a$ but in order of hours for another cascade $c_b$. Therefore, if we use the actual value of the infection times, we will not infer the links that $c_b$ spreads over the network; since the time differences in $c_b$ are much larger than $c_a$, the information obtained from $c_b$ would be ignored. To overcome this problem, we define a function that maps these continuous values to a set of discrete values and refer to the corresponding variable as \textit{infection index}. Here, our goal is to discover the links of a static network structure from the observed cascades. We do not consider the speed of cascades and expect a cascade to spread in the same manner as the other cascades in an identical path on the graph. We also ignore the impact of cascades on each other. Therefore, instead of the propagation time, the propagation sequence is important, and we achieve this representation by mapping time values to infection sequence order per each cascade. 

This way, each cascade $c_m$ is a vector containing pairs of nodes and infection indexes sorted in ascending order according to their infection indexes, as depicted in \equref{eq:cascadevector}. From now on, we call this vector the \textit{cascade vector}, $\tilde{c}_i$ where $r$ is the maximum index of non infinity infection time:

\begin{equation}\label{eq:cascadevector}
\tilde{c}_i=\{(v_0,l^{(i)}_{v_0}),(v_1,l^{(i)}_{v_1}),\ldots,(v_r,l^{(i)}_{v_r})\}
\end{equation}

\begin{figure*}[t]
	\centering
	\includegraphics[width=1\textwidth]{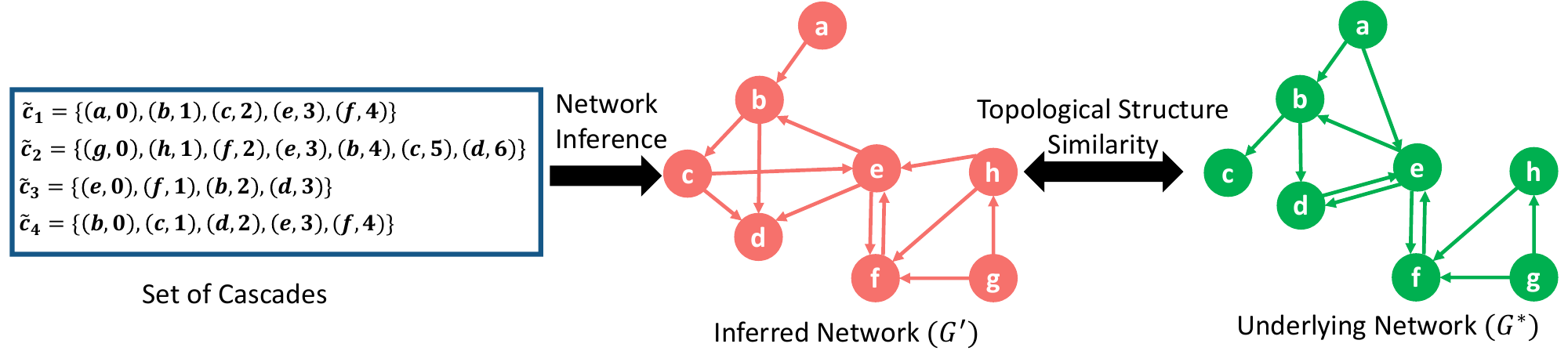}
	\caption{Problem definition. A set of cascades are the inputs. The proposed method should infer links between nodes. We evaluate the inferred network through topological similarities against ground truth.}
	\label{fig:ProblemDefinition}
\end{figure*}

As illustrated in \figref{fig:ProblemDefinition}, we assume that the underlying network $G^{*}= (V,E)$ is unknown, and the only observation is the set of cascades ($C$). If each network node participates in at least one cascade, the group of nodes $V$ can be extracted from the observation. We aim to obtain a weighted graph $G$ so that each weighted link ${0 \le w _{uv}}\le1$ is calculated considering the structural preservation properties. 
We would like to assign a weight to each possible edge of the complete graph ${G}$ with a weighted adjacency matrix $w$ that maximizes the following likelihood function:

\begin{equation}\label{eq:MLEG2}
{G'} = \argmax_{G|{0 \le w \le 1}}P(C|G)
\end{equation}

Then, we evaluate the similarity between the inferred network and the real underlying network based on the topological structure. The notations used in the paper are presented in \tabref{tab:notations}. 

\begin{table}[ht]
    \caption{Mathematical notations of this paper}
    \label{tab:notations}
    \small
    \centering
    \begin{tabularx}{\textwidth}{cX} 
        \hline\hline
        \textbf{Symbol} & \textbf{Definition} \\
        \hline\hline
        $N$, $M$ & Number of nodes and cascades \\
        $V$, $E$ & Set of nodes and directed edges in the underlying network \\
        $G^{*}$, $G$, $G'$ & Underlying, complete, and inferred network \\
        $(u,v)$ &  Direct link from node $u$ to node $v$\\
        $c_i$, $C$ & $i$'th cascade in set of cascades $C$ \\
        $t^{(i)}_u$ & Infection time of node $u$ in cascade $i$ \\
        $\tilde{c}_i$, $\tilde{C}$ & $i$'th ordered cascade in set of ordered cascades $\tilde{C}$ \\
        $l^{(i)}_u$ & Infection order of node $u$ in cascade $i$ \\
        $\lambda^{(i)}_{uv}$,$\lambda_{uv}$ & Diffusion Information behavior of nodes $u$ and $v$ in cascade $i$, in all cascades \\
        $In(u)$ & Set of cascades that $u$ has participated in \\
        $\theta_{uv}$ & Structural information behavior of nodes $u$ and $v$ \\
        $w_{uv}$ & Weight for edge from node $u$ to node $v$ in $G'$ \\
        $\overline{n_c}$ & Average number of nodes with non-infinity infection time in cascades  \\
        \hline\hline
    \end{tabularx}
\end{table}

\section{Our Method: DANI} \label{sec:DANI}

As mentioned before, our goal is to define a weight ${w_{uv}}$ for each edge $(u,v)$ that represents its existence probability in the underlying network. For this purpose, DANI first utilizes diffusion information and estimates the weight of all potential edges. In the second step, DANI assigns another weight to the joint behavior of nodes by considering the relation between structural properties and diffusion behavior. Finally, both weights are considered simultaneously to compute each edge's existing weight.

For simplicity, we assume the spread probability of cascades to be independent and identically distributed (i.i.d) over the graph. Therefore, the probability of observing the cascade set over the network is:
\begin{equation}\label{eq:CascadeOverNet}
P(C|G) = \prod\limits_{i = 1}^{i = M} {P({c_i}|G)} 
\end{equation}

Graph $G$ is composed of set of edges $(u,v)$, and we assume that all edges are mutually independent:

\begin{equation}\label{eq:GIndependence}
P(G) =\prod\limits_{(u,v) \in G} {P(u,v)}
\end{equation}

Therefore, the probability that cascades $c_i$ spreads over graph $G$ is given by \equref{eq:CiconditionalGTotal}, assuming that the joint probability of edges is conditionally independent:

\begin{equation}\label{eq:CiconditionalGTotal}
    \begin{split}
        P({c_i}|G) &= P({c_i}|{\{ (u,v)\} _{(u,v) \in E}}) \\
        &= \cfrac{{P({{\{ (u,v)\} }_{(u,v) \in E}}|{c_i}) \times P({c_i})}}{{P({{\{ (u,v)\} }_{(u,v) \in E}})}} \\
        &= \cfrac{{\prod\limits_{(u,v) \in E} {P((u,v)|{c_i})}  \times P({c_i})}}{{\prod\limits_{(u,v) \in E} {P(u,v)} }}
    \end{split}
\end{equation}

Where $P({c_i})$ is the prior probability for cascade formation, and we assume it is constant for all cascades. Therefore, we can approximate the probability of cascade $c_i$ propagating over $G$ by \equref{eq:CiconditionalG}:

\begin{eqnarray}\label{eq:CiconditionalG}
P({c_i}|G) \propto \prod\limits_{(u,v) \in G} {P({c_i}|(u,v))} 
\end{eqnarray}
Hence, by using \equref{eq:CiconditionalG} and applying log function, we can rewrite \equref{eq:CascadeOverNet} as:

\begin{eqnarray}\label{eq:LogCconditionalG}
\log(P(C|G))\propto \log\left[\prod \limits_{i = 1}^{i = M}{\prod \limits_{(u,v) \in G}{P\left(c_i|(u,v)\right)}}\right]
\propto \sum\limits_{(u,v) \in G} {\sum\limits_{{c_i} \in C} {\log (P({c_i}|(u,v))} }
\end{eqnarray}

where $P({c_i}|(u,v))$ represents the probability that cascade $c_i$ spreads through edge $(u,v)$. According to previous works \cite{Malmgren:2008,gray2020bayesian}, independent cascades over the network are Poisson point processes that follow the exponential distribution \cite{NetInf:2010}:

\begin{equation}\label{eq:CascadeProbability}
P({c_i}|(u,v)) = exp\left({\cfrac{{ -\lambda^{(i)}_{uv}}}{\theta_{uv} }}\right)
\end{equation}

Finally, our problem can be formalized as log-likelihood for all cascades over the network graph:

\begin{equation}\label{eq:FinalMLE}
G' = \argmax_{G}\sum\limits_{(u,v) \in G} {\sum\limits_{{c_i} \in C} {\log (P({c_i}|(u,v))}}
\end{equation}

We interpret $log(P(c_i|(u,v)))$ as the weight of cascade $c_i$ propagating on edge $(u,v)$, and call it ${w ^{(i)}_{uv}}$. 
We aim to estimate the weight of each presumable network link for each cascade $c_i$ and aggregate them to find the final weight of links. The rest of the paper introduces the DANI algorithm in which ${w_{uv}}$ is computed according to the relations between structural properties and diffusion information.

\textbf{Diffusion Information:} From each cascade vector $\tilde{c}_i$ we can construct a probable directed graph. Each node $u$ in this graph can get infected by a node $v$, if $l^{(i)}_v \prec l^{(i)}_u$. The probability of the existence of edge $(u,v)$ in this graph depends on the difference between the infection label of $u$ and $v$, such that more difference between the infection labels of these nodes in $\tilde{c}_i$, decreases the participation probability of link $(u,v)$ in the contagion \cite{NetInf:2010}:
\begin{equation}\label{eq:LinkParticipationProbability}
P(c_i|(u,v)) \propto \cfrac{1}{{l^{(i)}_v-l^{(i)}_u}}
\end{equation}

\begin{figure}[ht]
	\begin{center}
		\includegraphics[width=0.6\textwidth]{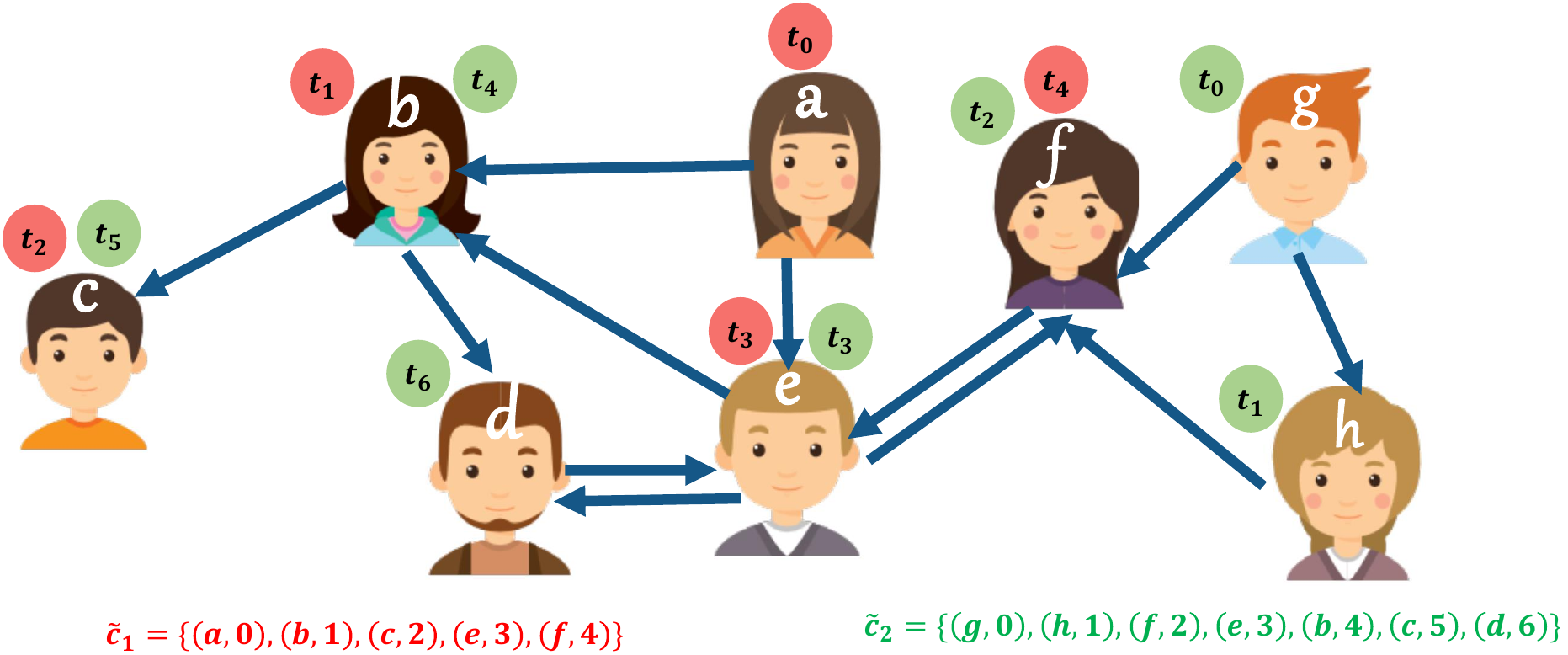}
		\includegraphics[width=0.5\textwidth]{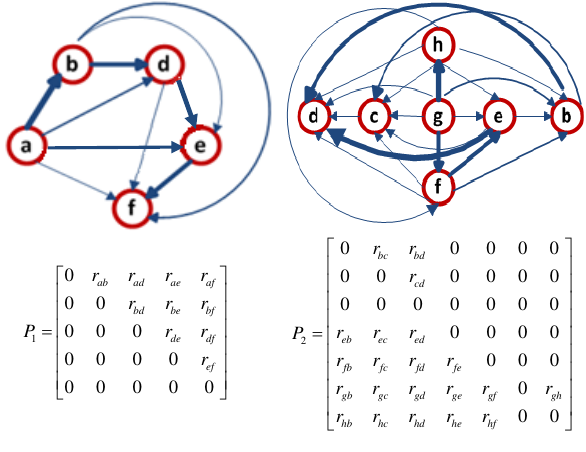}
		\caption{\label{fig:MarkovChainModel} (Up) A network topology and two cascades over the network. The infection of each user with contagion is represented by color with infection time. The cascade vector calculated by the proposed method is also considered. (Down) Markov chain and matrix for each cascade: probable directed graph with the DANI algorithm. Edge probabilities are marked with thick and thin lines. The Markov transition matrix per cascade is obtained from these graphs, as shown here.}
	\end{center}
\end{figure}

\figref{fig:MarkovChainModel}a illustrates a typical network and two of its distinct contagions. Their cascade vectors and corresponding probable graphs (simply referred to as graph) are shown in \figref{fig:MarkovChainModel}b and \figref{fig:MarkovChainModel}(c), respectively.

Each cascade vector $\tilde{c}_i$ can be modeled as a discrete-time stochastic process where nodes stand for independent states, and the sequence of nodes in the cascade vector $\tilde{c}_i$ represent the state of the process at discrete times. The transition probability between different states (nodes), or in other words, the contagion transmission probability from one node to another node, is equivalent to the weighted link between them. 

At any state of cascade, the contagion transmission probability of each node (state) depends on the likelihood of being at its neighboring node in the immediately preceding period. It is independent of its previous time states or the traversed nodes. This property is called the \textit{Markov Property}. Hence, it can be thought that the states of former times are all accounted for by incorporating the state of the last immediate time \textit{t}. Assuming the First Order Markov Property to hold, we have:

\begin{equation}
\begin{split}
P[X_{t + 1} = x_{t + 1} \rm {|} X_t = x_t,X_{t - 1} = x_{t - 1}, \ldots ,{X_0} = {x_0}]
= P[{X_{t + 1}} = {x_{t + 1}}\rm{|}{X_t} = {x_t}]
\end{split}
\end{equation}

The stochastic process that exhibits the Markov property is called a Markov Chain (MC). The number of states in our Markov chain is equal to the number of nodes, and each cascade vector $\tilde{c}_i$ is a finite Markov chain \cite{LevinPeresWilmer2006}. To define the Markov chain, we must compute its transition matrix. Each element of this matrix represents the probability of an edge in the network. As shown in \figref{fig:MarkovChainModel}, our goal is to find the transition matrix for each chain by utilizing the corresponding cascade information. 

At this point, we need to find the relation between the difference between the infection labels of two nodes and the probability that an edge exists between them. As discussed, this probability is inverse to the difference in their infection labels. Two approximations are used in the previous works:
\begin{enumerate}
	\item The probability of an edge only depends on the difference in the infection label of the nodes it connects. In other words, for all pairs of nodes with the same values of infection label difference, the existence probability is the same \cite{NetInf:2010}. 
	\item Consider a cascade vector such that a set of nodes ($D$) exists before $v$ in this vector. The probability that $u$ infected $v$ when $D =\{u\}$, is higher than the situation where $|D|\succ2$, because a node $z \in D$ with ${t_z} \prec {t_v}$ may have infected $v$. This will reduce the probability that $v$ has been infected by $u$. Moreover, the authors in \cite{EslamiRS11} proved that in addition to the difference in infection times, the number of infected nodes before node $v$, namely $|D|$, also affects the edge weights.
\end{enumerate}

Considering the above points, the existence probability of an edge between nodes $u$ and $v$ in the observed cascade vector $i$ is called $P(c_i|(u,v))$ which is inversely proportional to $l^{(i)}_v\times(l^{(i)}_v - l^{(i)}_u)$ where $l^{(i)}_v \succ l^{(i)}_u$.

Therefore, the Markov chain transition matrix $\lambda^{(i)}$, is a stochastic random matrix defined as:

\begin{equation}\label{eq:CascadeTransitionMatrix_D}
        d^{(i)}_{uv} ={{l^{(i)}_v\times(l^{(i)}_v - l^{(i)}_u})}
\end{equation}

\begin{equation}\label{eq:CascadeTransitionMatrix}
        {\lambda^{(i)}_{uv}} = \cfrac{{{d^{(i)}_{uv}}}}{{{ \sum \limits_{y \in V} {d^{(i)}_{uy}}}}}
\end{equation}

\textbf{Structural Information:} Modular structure of a network is one of the features of social networks, which has various effects on structural properties. As previously mentioned, community members have similar interests and behaviors, and diffusion is more likely to spread among them \cite{easley2010,Eftekhar:2013,BarbieriBM13,DBLP:conf/icdm/BarbieriBM13}. Intuitively, they appear in the same cascades more than nodes in different communities. Thus, the number of cascades passed between node $u$ and $v$ can be a sign of coherence between these nodes. If $u$ and $v$ are community members, they become infected jointly in more cascades compared to the cascades in which only one of them has been infected. We use the participations of pairs of nodes in different cascades as a mark for finding the similarity in their behaviors.

On the other hand, consider two boundary nodes that connect adjacent communities. These nodes may appear in the same cascades due to the bridge role of weak ties in the diffusion process.  However, the frequency of such cases is negligible compared to the number of cascades in which every single one has participated. 

According to the above justifications and the goal of inferring the links for each pair of node $(u,v)$, we define a function $\theta_{uv}$ that assigns a weight to edge $(u,v)$ by considering the nodes behavior based on the cascades information. In other words, this weight is a node-node similarity based on the structural properties that had impacted the diffusion information. The formula for $\theta_{uv}$ is similar to the Jaccard index \cite{jaccard}:
\begin{equation}\label{eq:edgeweight}
\theta_{uv} = \cfrac{{\left| {{In(u)\bigcap {In(v)} } } \right|}}{{\left| {{In(u)\bigcup {In(v)} }} \right|}}
\end{equation}
where $In(u)$ is the set of cascades that $u$ has participated in:
\begin{equation}
In(u) =
\left \{ {\begin{array}{*{20}{c}}
	{\bigcup\limits_{i =1}^M \{i\}} &{if(u \in \tilde{c}_i)}\\
	{ {\emptyset}} & {else}
	\end{array} } \right.
\end{equation}
For computing the numerator of \equref{eq:edgeweight}, $u$ should appear earlier than $v$ when both participate in each $\tilde{c}_i$.

The above node-node similarity can be used to detect the communities. The main idea of many structured-based community detection algorithms is grouping nodes based on the fact that when the similarity of two nodes is significant, they have more chance to be in a community \cite{du2007community,weng2013virality}. Since overlapping nodes are highly influential nodes \cite{6724361,ghorbani2016bayesian}, this formula also works for overlapping communities. Because the overlap nodes will take part in more common cascades and consequently gain large values.

\textbf{Total Information:} The probability of direct contagion propagation between two nodes would increase if the difference in their infection times decreases. In other words, the existence probability of edge $(u,v)$ is inverse to $\lambda_{uv}$. From a different perspective, larger values of $\theta_{uv}$ correspond to more probability that $u$ and $v$ are cohorts, indicating edge $(u,v)$ as an intra-community link. Therefore, the direct impact of $\theta_{uv}$ on the existence probability of edge $(u,v)$ will lead to infer the intra-community links that are the important links in the community detection step. 

By maximizing the logarithm of likelihood function \equref{eq:FinalMLE} using \equref{eq:CascadeProbability}, \equref{eq:CascadeTransitionMatrix}, \equref{eq:edgeweight} we obtain:
\begin{eqnarray}\label{eq:MinMaxLogMLEv1}
{G'}\propto \argmax_{G} ( \mathop \sum \limits_{\left( {u,v} \right) \in G} \mathop \sum \limits_{{c_i} \in C} ({\cfrac{{ -\lambda^{(i)}_{uv}}}{\theta_{uv} }}))
\end{eqnarray}

Since changing one of the fractions in \equref{eq:MinMaxLogMLEv1} does not change the others, we can write the MLE as:
\begin{eqnarray}\label{eq:MinMaxLogMLE}
{G'}\propto \argmax_{G} ( \mathop \sum \limits_{\left( {u,v} \right) \in G}  ({\cfrac{\theta_{uv} }{{ \lambda_{uv}}}}))
\end{eqnarray}

Where $\lambda_{uv}$ is the sum of the transition matrix of all cascades, preserving the Markov matrix properties, and it stands for the total transition matrix of the Markov chain. We name ${w_{uv}={\cfrac{\theta_{uv} }{{ \lambda_{uv}}}}}$ as the weight of link.

\SetKwComment{Comment}{/* }{ */}
\RestyleAlgo{ruled}

\begin{algorithm}[!b]
    \small
    \setstretch{1.3}
	\caption{DANI Pseudo-Code}
	\label{alg:DANI}
	\KwIn{Set of cascades over network ($C = \left\{ {{c_1},{c_2}, \ldots {c_M}} \right\}$)

}
	\KwOut{Inferred network ($G'$)}
	\For{each ${c_m} = \left\{ {\left( {{v_i},{t^{(m)}_i}} \right)} \right\} \in C$}
	{
		$l^{(m)}_{v_i} = index(t^{(m)}_{i})$\;
		$\tilde{c}_{m} = \left\{ {\left( {{v_i},l^{(m)}_{v_i}} \right)} \right\}$\;
		\For{each node $u \in \tilde{c}_m , v \in \tilde{c}_m$}
		{
			\If{($l^{(m)}_{u} < l^{(m)}_{v}$)}
			{
				\textbf{Compute} ${d^{(i)}_{uv}}$ according to \equref{eq:CascadeTransitionMatrix_D}\;
			}
		}
		\For{each $\left( {u,v} \right) \in {d^{(i)}}$}{
			$\lambda^{(i)}_{uv} \leftarrow \cfrac{{{d^{(i)}_{uv}}}}{{\mathop \sum \nolimits_{y \in V} {d^{(i)}_{uy}}}}$\;
			$\lambda^{t}_{uv} \leftarrow \lambda^{t}_{uv} + \lambda^{(i)}_{uv}$\;
		}
	}
	\For{each $\left( {u,v} \right) \in \lambda^{t}$}
	{$\lambda^{t}_{uv} \leftarrow \cfrac{\lambda^{t}_{uv}}
			{{\sum\limits_{y \in V} {\lambda^{t}_{uy}} }}$ \;
		$\lambda_{uv} \leftarrow \lambda_{uv} + \lambda^{t}_{uv}$\;
	}
	\For{each $(u, v) \in {\lambda}$}{
		\textbf{Compute} $\theta_{uv}$ according to \equref{eq:edgeweight}\;
		$w_{uv} = \cfrac{\theta_{uv}}{\lambda_{uv}}$;}
	
	\For{{$E_{ij}$} $\in$ {$G$}}{
			\If{($w_{ij} > 0$)}
			{
	                {$G'$} $\leftarrow$ ${G'}$ $\bigcup {\{ {E_{ij}} \} }$ \;
			}
	}
	\textbf{Return} {$G'$}\;
\end{algorithm}

The DANI pseudo-code is presented in \algref{alg:DANI}, where $\lambda^{t}$ is used to keep the sum of all transition matrices of cascade vectors.  Then, $\lambda^{t}$ is normalized to obtain the stochastic transition matrix $\lambda$ based on the diffusion information. In the next step, by considering the structural properties, $\theta_{uv}$ is calculated and used along with $\lambda$ to produce the inferred weighted edges of network $w$. The normalization in the process of computing $\lambda$ keeps its stochastic transition property and its value within the limits of $\theta_{uv}$ values. Hence, their multiplication stays in a fixed range for all the possible edges. 

\subsection{Theoretical Analysis of DANI}

Consider a cascade $c_m$ with $n_{c_m}$ infected nodes.
According to \algref{alg:DANI}, there are $\binom{n_{c_m}}{2}$ pair of nodes $(u,v)$ for calculating $d_{uv}$ and other normalizations with the time complexity of $O(n^2_{c_m})$. Also, $In(u)$ should be considered for the set of nodes. Therefore, having $M$ different cascades with an average number of nodes $\overline {{n_c}}$ and the total number of participated nodes $N$ as the input, the DANI algorithm has a complexity of $O(M \times {{\overline {{n_c}}}^2 \times N)}$. 

\begin{table}[!b]
    \caption{Time complexity of DANI and other methods. $\kappa$, $\eta$, and $d$ are the dimension of embedding, number of iterations, and depth of neural network, respectively.}
    \label{tab:timecomplexity}
    \small
    \centering
    \begin{tabularx}{0.6\textwidth}{YY} 
        \hline\hline
        \textbf{Method} & \textbf{Time Complexity} \\
        \hline\hline
        DANI & $O(MN{\overline{{n_c}}}^2)$ \\
        NetInf & $-$ \\
        DNE & $O(MN^2)$ \\
        MultiTree & $-$ \\
        NetRate & $-$ \\
        KEBC & $O(MN^2\kappa^3)$ \\
        REFINE & $O(N^2M+N^3+N\kappa^{2d}\eta)$ \\
        PBI & $O(MN^2\eta)$ \\
        \hline\hline
    \end{tabularx}
\end{table}

Some baseline studies did not provide time complexity analysis for their algorithms because their methods are heavily dependent on network structure \cite{NetInf:2010}. MultiTree and NetInf use greedy hill climbing to maximize the corresponding sub-modular function through an iterative approach where an edge is inferred at each step if a marginal improvement is achieved by adding the edge. Solving the convex objective function of Netrate is time consuming.
On the other hand, DANI is free from iterations and thus outperforms related works in term of running time. An overview of the runtime complexity of related methods can be found in \tabref{tab:timecomplexity}. Consequently, DANI has a significant advantage over the existing methods in terms of running time.

\subsection{Distributed pipeline}

\begin{figure}[t]
    \centering
    \includegraphics[width=0.8\textwidth]{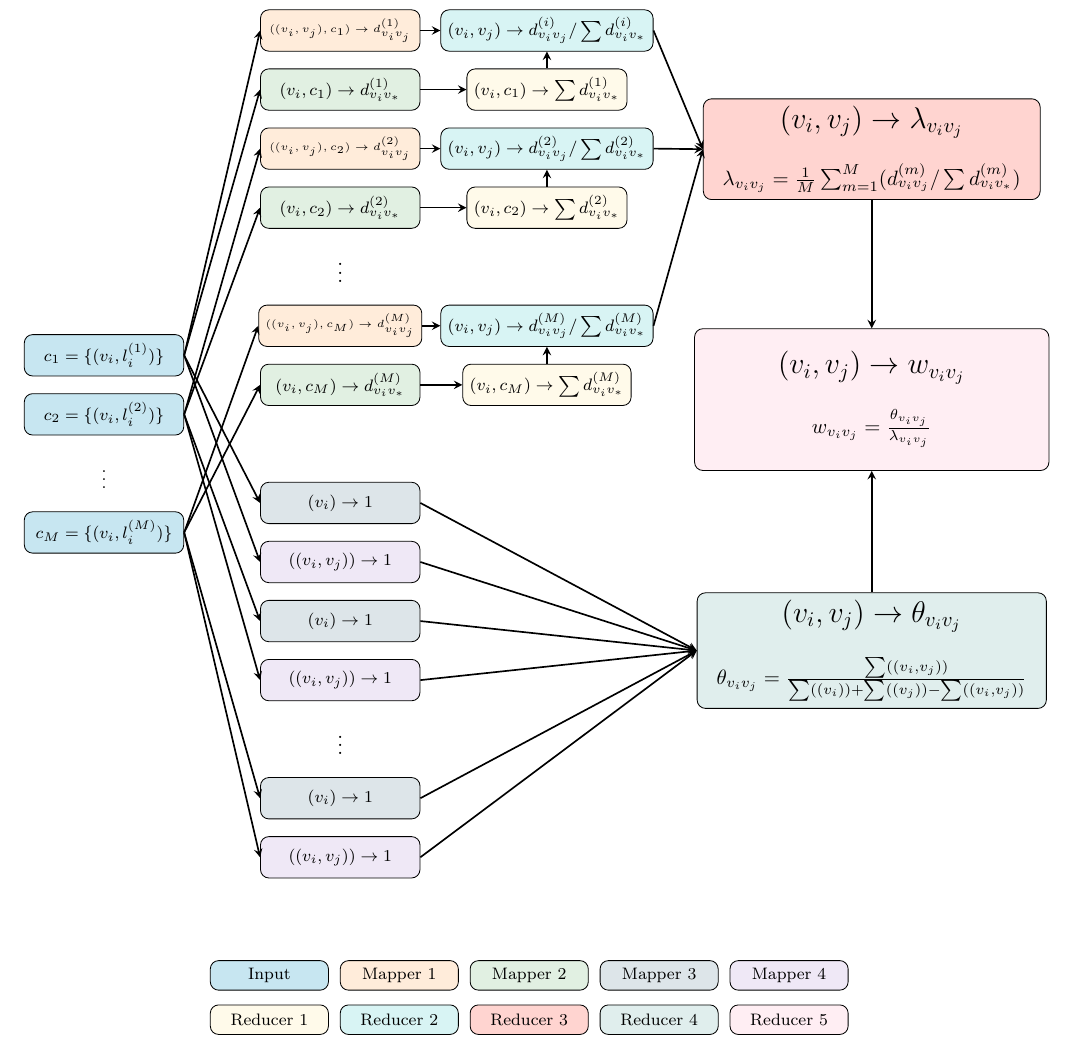}
    \caption{DANI Algrotithm in the MapReduce framework}
    \label{fig:mapreduce}
\end{figure}

It has already been mentioned that the time complexity of DANI depends on the number of users involved in cascades (nodes of the network) and the number of cascades, which could be quite large. While DANI has low order complexity compared to previous works, parallel methods such as MapReduce can increase its scalability when applied to massive data. MapReduce pattern is a distributed data processing framework that works mainly through Map and Reduce functions \cite{mapreduce2015}. The Map function produces key $\rightarrow$ value pairs in parallel. Then the same keys are grouped, and the Reduce function is applied in parallel to agglomerate values of the group for a key. The $M$ cascades are stored on HDFS in an index-based format known as $\tilde{c}_i$. The data is split across multiple machines. As shown in \figref{fig:mapreduce}, each Mapper1 receives cascades and maps each pair of nodes $h_{ij}=(v_i,v_j)$ in cascade $m$ with $l^{(m)}_{i}<l^{(m)}_{j}$ and the cascade name $m$ to the value from \equref{eq:CascadeTransitionMatrix_D}. This equation is also the value for the key in Mapper2, which contains the first node of $h_{[ij]}$. In Reducer1, the sum operation aggregates all outputs of Mapper2 based on the same key. Now the Markov chain transition matrix $\lambda^{(m)}$ of \eqref{eq:CascadeTransitionMatrix} can be calculated for each pair of nodes in Reducer2. $\lambda_{v_{i}v_{j}}$ is the output of Reducer3 grouped according to keys having pairs of nodes. Mapper3 maps each node in a cascade to value $1$, while Mapper4 maps each pair of nodes in a cascade with $l^{(m)}_{i}<l^{(m)}_{j}$ to value $1$. By grouping the output of these mappers, using sum aggregate and \eqref{eq:edgeweight}, $\theta_{v_iv_j}$ is produced in Reducer4. At the end of the process, Reducer5 divides the values of corresponding keys of Reducer3 and Reducer4 and outputs $w_{v_iv_j}$.

\section{Experiments}\label{sec:experiments}

The following section experimentally analysis DANI and compares our method's empirical results with related works. Our analysis focused on two different aspects: (1) inferring the underlying network correctly and (2) preserving topological structures. We used both real and synthetic datasets to evaluate models' performance.

We have compared our method with other works which infer a static network under the same conditions, including: \textbf{(1) NetInf} and \textbf{(2) MultiTree} from the category of tree-based-maximum-likelihood-estimation methods, \textbf{(3) NetRate} from the category of convex-based-maximum-likelihood-estimation methods, \textbf{(4) DNE} as a Markov-random-walk-based method, \textbf{(5) PBI} which is a Bayesian-inference-based method, \textbf{(6) KEBC} and \textbf{(7) REFINE} which are embedding-and-clustering-based methods.

We have implemented our method in both C++ and Python\footnote{https://github.com/AryanAhadinia/DANI}, and both versions are available on GitHub.
For competing works, the official code released by the authors is used, except REFINE, which is not available. REFINE code has been implemented and its correctness verified with the paper's results. In an unfair manner, different hyper-parameters of REFINE are tuned on the underlying network, and the best results are reported. The experiments were conducted on a server running Ubuntu 18.04 with an Intel® Core™ i9-7980XE processor and 128GB of memory. Additionally, we used NVIDIA GeForce GTX 1080TI GPUs to train deep neural networks.

\subsection{Metrics}\label{sec:metric}

The metrics used to evaluate the methods can be divided into three main categories: (1) metrics which are evaluating how good the model is in inferring links will be explained in subsection in \ref{subsec:metrics_1}, (2) metrics which are assessing how similar the structure of the inferred network is to the underlying one will be explained in subsection in \ref{subsec:metrics_2}, and (3) metrics measuring structural characteristics of a network will be explained in subsection in \ref{subsec:metrics_3} which we can use to compare the underlying and the inferred network characteristics.

For measuring how good the models perform in structure preservation, the concept of \textit{community} is important, a tangible concept in sociology and social networks. An algorithm is utilized for both ground truth and inferred networks to detect communities. We reviewed several community detection algorithms to find one that works with a variety of networks, including directed and undirected networks, detects communities with high accuracy, requires no prior knowledge of the number of output communities, and requires minimal memory and processing time for a sparse graph.
Finally, the \textit{OSLOM} algorithm proposed by \cite{lancichinetti2011finding} was chosen which meets the above requirements. OSLOM is the first algorithm that uses statistical features to identify communities. To assess accuracy, OSLOM uses local modularity measurement to avoid global errors.

\subsubsection{Correctness}\label{subsec:metrics_1}

\begin{itemize}
    \item \textit{F1-score (F1)}: We have used the harmonic mean of precision and recall, named F1-score, as a classic frequently used metric for classification problems. 

    \item \textit{Macro F1 (MF1) and Micro F1 (mF1)}: Link prediction is a highly class imbalance problem since most of the possible links are not formed in a social graph. Therefore, we use Macro and Micro F1, which are metrics for class imbalance problems. Micro and Macro F1 are the averages of F1 in the level of samples and classes, respectively.
\end{itemize}

\subsubsection{Structural similarity preservation}\label{subsec:metrics_2}

\begin{itemize}

    \item \textit{Jensen-Shannon distance (JS)}: The Jensen-Shannon distance is an extension of the Kullback–Leibler divergence, which measures the distance between two probabilistic distributions symmetrically. We use the Jensen-Shannon distance to measure the distance between the distribution of nodes degrees in underlying and inferred networks. Jensen-Shannon distance is defined in \equref{eq:js}. 
    
    \begin{equation}\label{eq:js}
        JS(P, Q) = \sqrt{\frac{KL(P || Q) + KL(Q || P)}{2}}
    \end{equation}
Where $P$ and $Q$ are distribution of nodes degrees for underling and inferred networks, respectively. The value of \textbf{JS} distance is greater or equal to zero. Zero indicates that $P$ and $Q$ are completely similar to each other. Higher value of \textbf{JS} indicates more difference between two distributions. 

    \item \textit{Normalized mutual information (NMI)}: NMI is a probabilistic information theoretical metric proposed by \cite{manning2008introduction} that compares the similarity of two sets of clusters. This metric is defined in \equref{eq:nmi}.

    \begin{equation}\label{eq:nmi}
        NMI(C_A, C_B) = \frac{-2\sum_{i=1}^{|C_A|}\sum_{j=1}^{|C_B|}N_{ij}\log\left(\frac{N_{ij}N_{..}}{N_{i.}N_{.j}}\right)}{\sum_{i=1}^{|C_A|}N_{i.}\log \left(\frac{N_{i.}}{N_{..}}\right) + \sum_{j=1}^{|C_B|}N_{.j}\log \left(\frac{N_{.j}}{N_{..}}\right)}
    \end{equation}

    Where $C_A$ and $C_B$ are two sets of clusters. $N$ is a $|C_A|\times |C_B|$ matrix in which $N_{ij}$ is equal to the number of common members of $i$'th cluster of $C_A$ and $j$'th cluster of $C_B$. $N_{i.}$, $N_{.j}$, and $N_{..}$ are equal to the sum of $i$'th row, $j$'th column, and the whole of the matrix N, respectively. The range of NMI values is between $0$ and $1$, where NMI of $1$ represents the exact matching of two sets, and $0$ shows the maximal difference.
    
    \item \textit{Pairwise F1-score (PWF)}: Consider a pair of nodes $u$ and $v$. These two nodes can be in same or different community in the ground truth and the inferred network. Consider the classification problem where we want to determine whether nodes $u$ and $v$ are in the same community or not.
    Formally, let $H_{G^*}$ and $H_{G'}$ represent the sets of node pairs in the same community at the underlying and the inferred network, respectively. PWF is defined in \equref{eq:pwf} \cite{yang2009bayesian,conf/icde/QiAH12}.
    
    \begin{equation*}
        \begin{split}
            PW_{precision} &= \cfrac{{|{H_G} \cap {H_{G^*}}|}}{{|{H_{G'}}|}} \\
            PW_{recall} &= \cfrac{{|{H_G} \cap {H_{G^*}}|}}{{|{H_{G*}}|}}
        \end{split}
    \end{equation*}
    
    \begin{equation}\label{eq:pwf}
         PWF = \frac{2 PW_{precision} PW_{recall}}{PW_{precision} + PW_{recall}}
    \end{equation}
    
    We considered PWF, because it is not enough to rely solely on NMI \cite{journals/abs-1110-5813}. In addition, because of sparsity in real networks, the obtained modularity values are often incorrect. 
\end{itemize}

\subsubsection{Metrics for evaluating networks characteristics}\label{subsec:metrics_3}

In this section, we talk about metrics that measure the structural characteristics of a network. These metrics give us two separate values for the ground truth and the inferred network. We use their relative difference ($\Delta$) to compare these two values, as defined in \equref{eq:delta}.

\begin{equation}\label{eq:delta}
    \Delta\xi = \frac{|\xi_{gt}-\xi_{in}|}{\xi_{gt}}
\end{equation}

Where $\xi_{gt}$ and $\xi_{in}$ are the values of the measured metric for ground truth and inferred network, respectively, and $\Delta\xi$ in the relative difference for metric $\xi$.

\begin{itemize}
    \item \textit{Average community size (ACS)}: The average community size is utilized as a metric to determine whether the inferred network has the same community distribution as the underlying network \cite{garrels2023lazyfox}.
    
    \item \textit{Conductance (Cnd)}: For each cluster, conductance is defined as the ratio of the number of inter-cluster links connected to a node in that cluster to the number of all links connected to at least one node in that cluster.
    
    \begin{equation}\label{eq:Conductance}
        C_G(S) = \frac{c_G(S)}{\sum_{u \in S}D_G(u)}
    \end{equation}
    
    In \equref{eq:Conductance}, $c_G(S)$ is the number of inter-cluster links which one of its ends is a node in cluster $S$ and $D_G(u)$ is the degree of $u$ in graph $G$. For a network, conductance is the average conductance for each cluster.
        
    \item \textit{Number of connected nodes (CN)}: The previous works only focused on the number of correct inferred links. Despite their high accuracy, no connected links may be detected at some nodes. Hence, these nodes are omitted in the inferred network and consequently would collapse the topological properties. As a result, only a fraction of the underlying nodes, presented in the inferred network, is used in evaluations.
    
    \item \textit{Density ($\rho$)}: For each cluster, density is calculated as the ratio of the number of links in a cluster to the number of all possible links in that cluster. The overall density of a network is calculated by averaging the density of each of its clusters.
    Density and conductance are complementary measures and are often considered together.
    For instance, the density of a k-clique subgraph with many links to the rest of graph is high, while its conductance low. Therefore, these two criteria are often used together to evaluate the candid methods.
    In this paper, we desire to have more similarities in density and conductance among original and inferred networks.
        
    \item \textit{Average Clustering Coefficient (CC)}: For a node $u$ in an undirected network $G$, the clustering coefficient is the ratio of the number of triangles formed around $u$ to the total number of triangles that can be formed with edges connected to $u$. Intuitively, this measure wants to discover how likely it is that $v$ and $t$ are also connected if $u$ is connected to both $v$ and $t$. The clustering coefficient is defined in \equref{eq:cc}.
    
    \begin{equation}\label{eq:cc}
        CC_G(u) = \frac{|\{\{v, t\} | \{u, v\} \in E_G \land \{u, t\} \in E_G\}|}{\binom{D_G(u)}{2}}
    \end{equation}
    
    Where $E_G$ is the set of undirected edges in graph $G$, and each pair of $\{s, w\}$ is an undirected edge in this set. $D_G(u)$ is the degree of $u$ in graph $G$.
    
    The average Clustering Coefficient of a graph is the average of clustering coefficients for all nodes in the graph.
\end{itemize}

For better readability of the article, the abbreviations of these metrics are given in the \tabref{tab:abbreviation}

\begin{table}[t]
    \caption{Metrics abbreviation of this paper}
    \label{tab:abbreviation}
    \small
    \centering
    \begin{tabularx}{\textwidth}{cX} 
        \hline\hline
        \textbf{Abbreviation} & \textbf{Metric} \\
        \hline\hline
        \textbf{F1}, \textbf{MF1}, \textbf{mF1} & F1-score, Macro F1 , and Micro F1 \\
        \textbf{JS}  & Jensen-Shannon distance for degree distribution between inferred and underlying networks\\
        \textbf{NMI} & Normalized mutual information\\
        \textbf{PWF} & Pairwise F1-score\\
        $\boldsymbol{\Delta}$ & Relative difference\\
        \textbf{ACS}  & Average Community Size\\
        \textbf{Cnd} & Conductance\\
        \textbf{CN} & Number of connected nodes\\
        $\boldsymbol{\rho}$ & Density\\
        \textbf{CC} & Average Clustering Coefficient\\
        \hline\hline
    \end{tabularx}
\end{table}

\subsection{Datasets}\label{sec:datasets}

\subsubsection{Synthetic datasets}

We used the LFR benchmark proposed in \cite{PhysRevE.80.016118}, which provides actual network characteristics, and its properties can be controlled to provide networks with built-in communities. We utilize parameter $\mu$ to control how community-structured the network is. For each node, $\mu$ represents the number of connected inter-community links to the number of all connected links ratio. The value of $\mu$ can change between $0$ and $1$, and its alterations would change the network's community structure. Larger values of $\mu$ correspond to less community structure in the network while decreasing this parameter would improve the network's community structure. We have generated five networks with $1000$ nodes, average degree $15$, max degree $50$, minus exponent for the degree sequence $2$, minus exponent for the community size distribution $1$, min community sizes $20$, max community sizes $50$, and the value of $\mu$ parameters equals $0.1$, $0.2$, $0.3$, $0.4$, and $0.5$. According to the experimental results, values over $0.5$ for $\mu$ are unsuitable for highlighting community structures. We generated various artificial cascades for each of these networks by using the SNAP library proposed by \cite{SNAP} with different models. On each LFR-synthesised network, we have generated nine sets of cascades with a cardinality of $500$, $1000$, $3000$, $4000$, $5000$, $7000$, $10000$, $12000$, and $15000$, so we have a total of 45 different datasets, which enable us to study the effect of dataset characteristics on inference performance. The structure properties of the five structure networks of these datasets in listed in \tabref{tab:lfr_chars}. Column $\boldsymbol{\overline{n_c}}$ is the average and variance for length of cascades in $9$ different set of diffusion constructed over the basic structure network.

\begin{table}[H]
    \caption{Characteristics of LFR datasets}
    \label{tab:lfr_chars}
    \small
    \centering
    \begin{tabularx}{\textwidth}{YYYYYYY} 
        \hline\hline
        \textbf{Name} & \textbf{$\boldsymbol{\mu}$}  & \textbf{ACS} & \textbf{Cnd} & \textbf{CC}    & $\boldsymbol{\rho}$ & $\boldsymbol{\overline{n_c}}$  \\
        \hline\hline
        LFR1 & 0.1 & 35.71             & 0.056       & 0.267 & 0.198 & $31.0\pm0.2$  \\ 
        \hline
        LFR2 & 0.2 & 35.71             & 0.119       & 0.190 & 0.177 & $46.8\pm0.3$  \\ 
        \hline
        LFR3 & 0.3 & 29.41             & 0.189       & 0.158 & 0.194 & $64.9\pm0.3$  \\ 
        \hline
        LFR4 & 0.4 & 35.71             & 0.253       & 0.086 & 0.141 & $69.3\pm0.9$  \\ 
        \hline
        LFR5 & 0.5 & 35.71             & 0.340       & 0.051 & 0.117 & $76.6\pm0.3$  \\
        \hline\hline
    \end{tabularx}
\end{table}

\subsubsection{Real datasets}

In this paper, we refer to a dataset as real, when its topology and corresponding diffusion are accessible. We utilized five real datasets to evaluate the model's performance on real-world datasets and to evaluate the scalability of the model: The MemeTracker project collected data from March 2011 to February 2012 in which nodes are online news websites, cascades are meme propagation, and links are hyperlinks between websites \cite{MemeTracker}. A sample dataset from this project was used here as the MemeTracker dataset, and three datasets were constructed by filtering the total memes with keywords related to LinkedIn, NBA, and News of the World. Also, we utilized a large Twitter network that sampled public tweets from the Twitter streaming API in the range of 24 March to 25 April at 2012 \cite{weng2013virality}. By utilizing the follower network and hashtag retweeting, we did some preprocessing by: 1) Considering any unique pair of hashtag and source ID as a separate cascade, 2) Keeping links that were being involved in at least one cascade, 3) Choosing users taking part in more than 2 cascades, and 4) Removing nodes of cascade thfat were not a member of network nodes. The characteristics of these datasets are listed in \tabref{tab:real_chars}.

\begin{table}[H]
\centering
\small
\caption{Characteristics of real datasets}
\label{tab:real_chars}
\begin{tabularx}{\textwidth}{YYYYYYYYYY}
\hline\hline
\textbf{Name} & \textbf{\#Nodes}  & \textbf{\#Links}   & \textbf{\#Cascades}    & $\boldsymbol{\overline{n_c}}$ & \textbf{\textbf{ACS}} & \textbf{Cnd} & \textbf{CC} & $\boldsymbol{\rho}$  \\ \hline\hline
LinkedIn      & 1035   & 9215167   & 22963     & 2.95     & 13.10                 & 0.383         & 0.272       & 0.214             \\ \hline
MemeTracker   & 4452   & 54447     & 1598225   & 4.68       & 62.84                 & 0.871         & 0.584       & 0.523             \\ \hline
NBA           & 2056   & 11955546  & 142737    & 2.90      & 24.18                 & 0.402         & 0.318       & 0.306             \\ \hline
Twitter       & 124023 & 213875    & 182783    & 2.29      & 5.37                  & 0.479         & 0.031       & 0.089             \\ \hline
World         & 1390   & 10546071  & 64618     & 2.93     & 21.71                 & 0.347         & 0.335       & 0.298             \\ \hline\hline
\end{tabularx}
\end{table}

\subsection{Experimental Analysis of DANI}\label{sec:ExAnalysis}

\begin{figure}[t]
    \centering
    \begin{subfigure}{0.495\textwidth}
        \includegraphics[width=\textwidth]{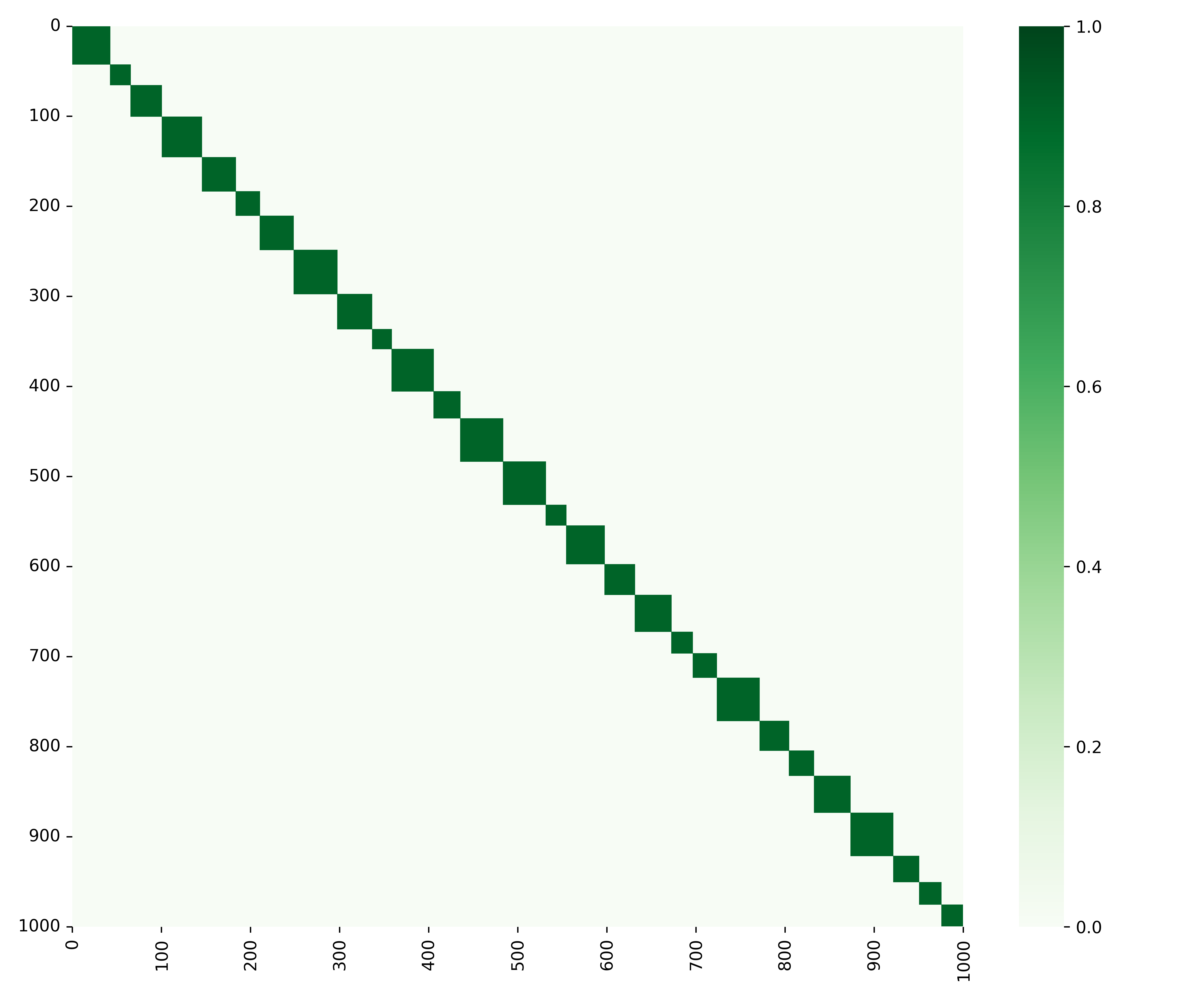}
        \caption{Ground truth}
        \label{fig:CommunitySimilarityGT}
    \end{subfigure}
    \hfill
    \begin{subfigure}{0.495\textwidth}
        \includegraphics[width=\textwidth]{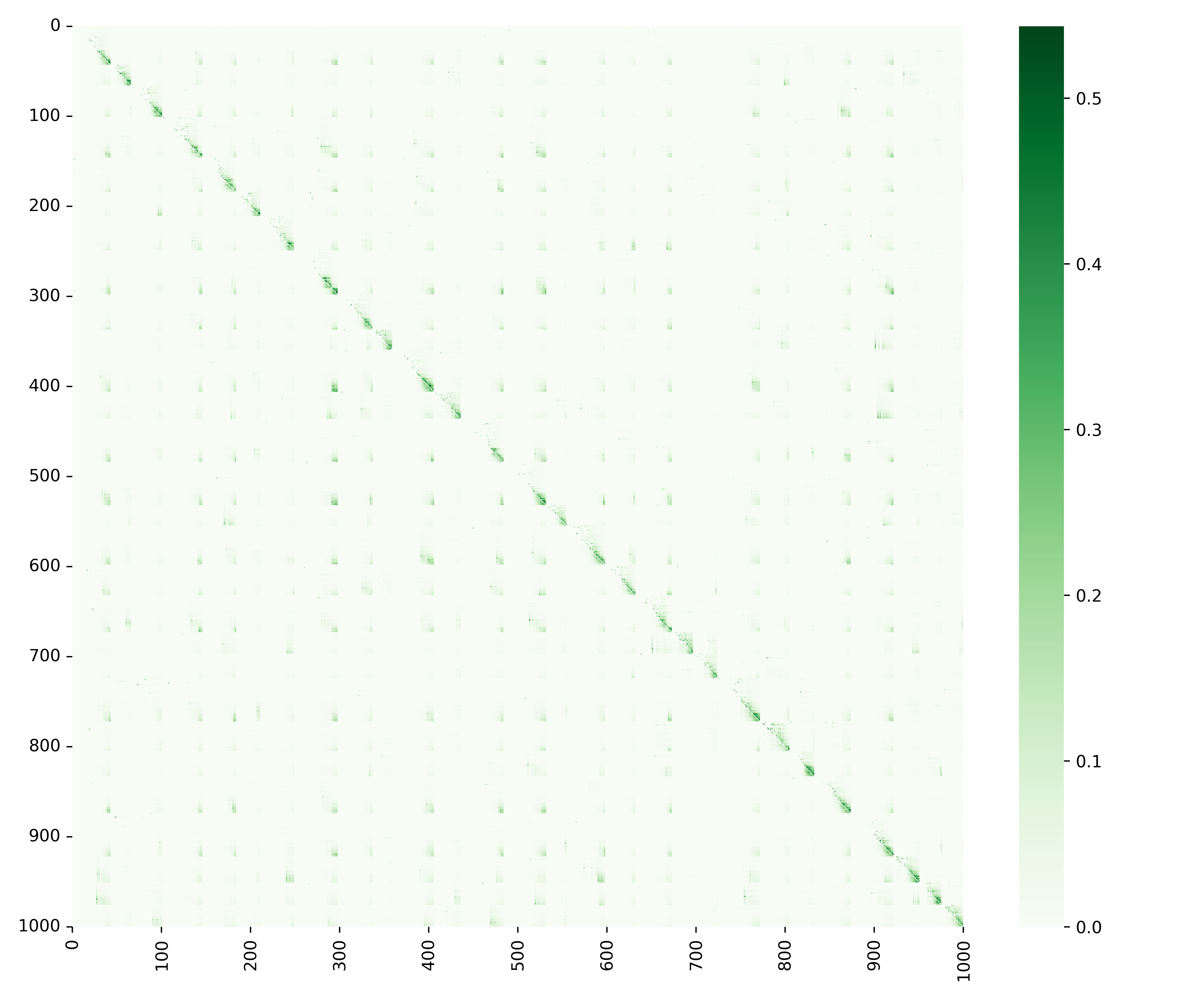}
        \caption{\equref{eq:edgeweight} value for each nodes pair}
        \label{fig:CommunitySimilarityDANI}
    \end{subfigure}
    \caption{The heat map indicates the score assigned to each edge. \figref{fig:CommunitySimilarityGT} is for the ground truth network in which all edges are deterministically labeled. Cells between $i$ and $j$ are hot if there is an edge between $i$ and $j$ and are cool if there is no edge. The nodes are rearranged so that each community forms a block. Each cell in \figref{fig:CommunitySimilarityDANI} shows the value of \equref{eq:edgeweight} for each pair of nodes $(i, j)$. Darker cells indicate higher values which show higher diffusion-based similarity between two nodes. This leads to the conclusion that communities can be specified using node-node diffusion-based similarity.}
    \label{fig:CommunitySimilarity}
\end{figure}

We provided a data-driven study to show the relation between \equref{eq:edgeweight} and the abovementioned concept of node-node similarity. To this end, the LFR1 dataset with 1000 nodes, 7692 links, 28 communities, and 10000 cascades is utilized. We changed the name of nodes such that nodes of a community are consecutively listed. In \figref{fig:CommunitySimilarity}, each cell $i, j$ in the heat map \figref{fig:CommunitySimilarityGT} indicate if there is an edge between nodes $i$ and $j$ and each cell $i, j$ in the heat map \figref{fig:CommunitySimilarityDANI} indicates the values of $\theta_{ij}$. We have used the Student's \textit{t}-test to show that the score assigned by \equref{eq:edgeweight} to intra-community links is significantly higher than the score assigned to extra-community links. The p-value calculated by \textit{t}-test is $0$, showing a significant difference between the means of scores assigned to intra-community and inter-community links.

One of the most important features of our proposed method is its high execution speed. The primary implementation of DANI is in C++, which brings high speed of execution due to direct execution on the hardware. Note that the execution speed of a program can be highly variable depending on the implementation language and details. Since we used official implementations provided by authors of competing works that are implemented in different languages, comparing their execution times is unfair. For this reason, we only deal with the theoretical and experimental analysis of DANI's execution time.

\begin{figure}[H]
    \centering
    \begin{subfigure}{0.495\textwidth}
        \includegraphics[width=\textwidth]{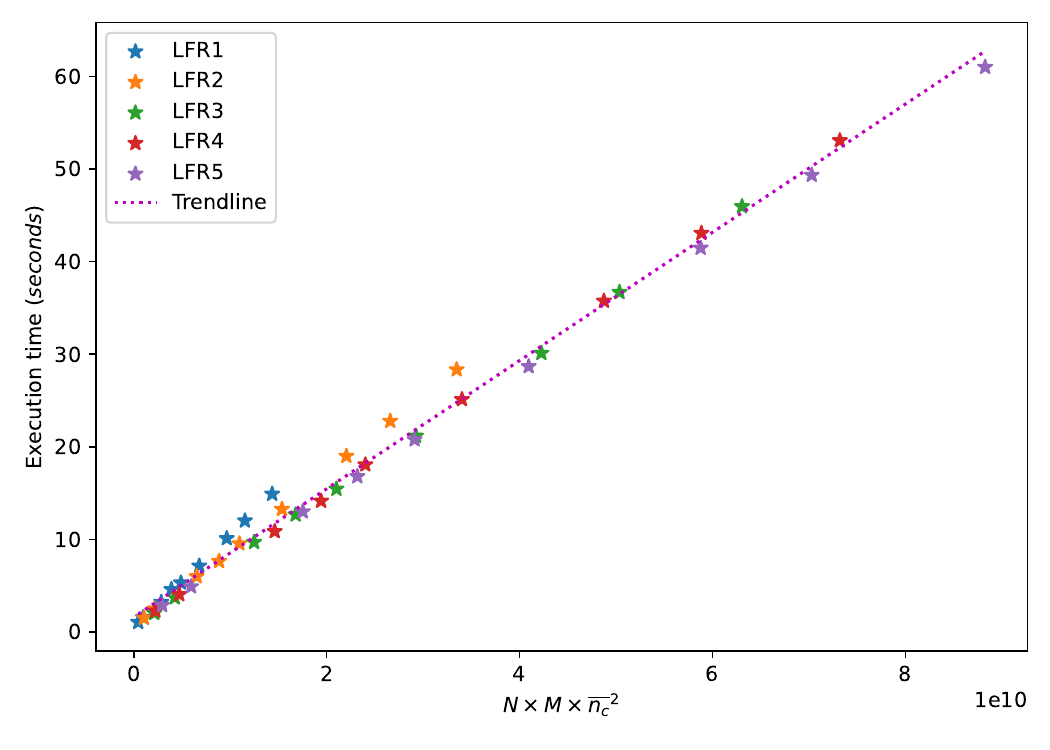}
        \caption{LFR Datasets}
        \label{fig:time_comp_lfr}
    \end{subfigure}
    \begin{subfigure}{0.495\textwidth}
        \includegraphics[width=\textwidth]{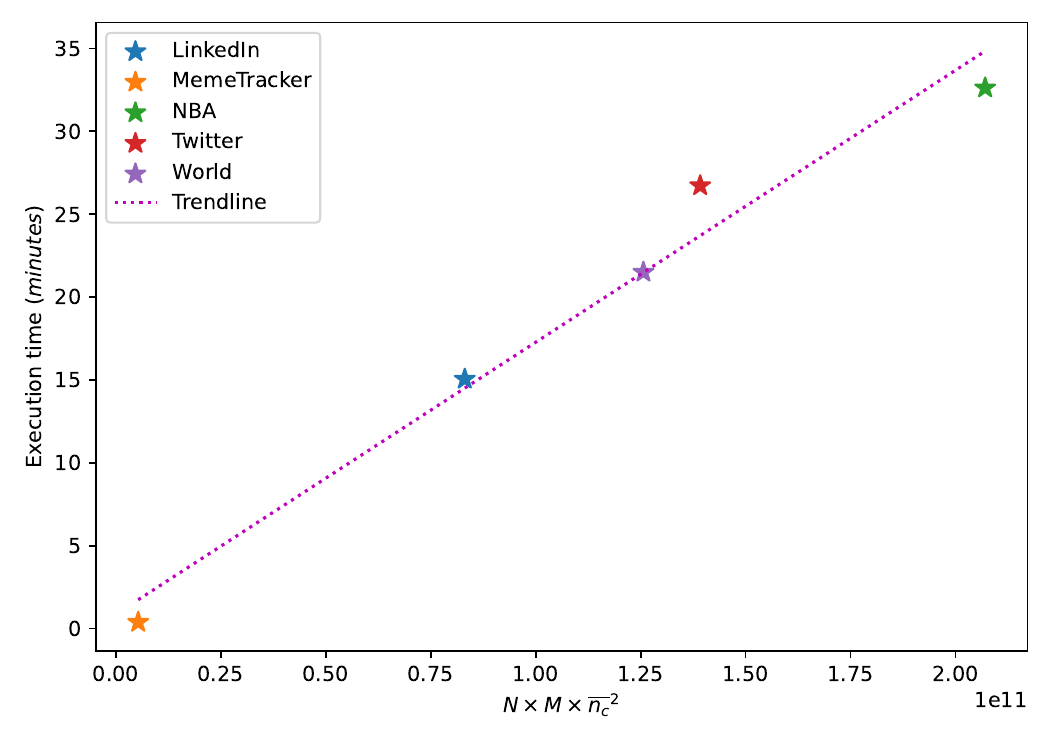}
        \caption{Real Datasets}
        \label{fig:time_comp_real}
    \end{subfigure}
    \caption{Execution time of DANI. \figref{fig:time_comp_lfr} indicates execution time on LFR datasets. Note that for each LFR dataset, we have generated nine sets of cascades, so we have a total of 45 values as execution time for each LFR dataset with each set of cascades. \figref{fig:time_comp_real} indicates execution time on real datasets.}
    \label{fig:time_complexity}
\end{figure}

Previously, we discussed DANI's time complexity and showed that the DANI algorithm can be executed in $O(MN\overline{n_c}^2)$ time which is linear with respect to both $M$ and $N$. In \figref{fig:time_complexity}, we record the execution time of DANI on all of our datasets to empirically prove that DANI can run at the time we have theoretically proven.

\begin{figure}[H]
    \centering
    \includegraphics[width=0.5\textwidth]{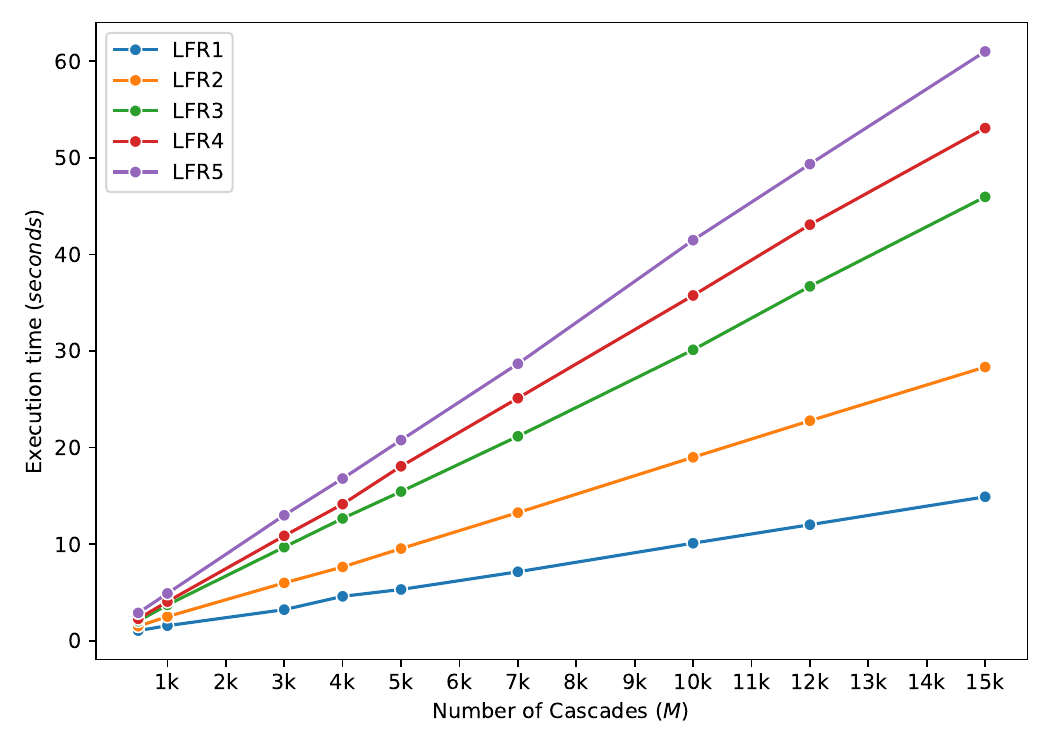}
    \caption{Execution time of DANI on LFR datasets with different numbers of cascades and value of $\mu$}
    \label{fig:time_lfr}
\end{figure}

As illustrated in \figref{fig:time_lfr}, the execution time of DANI is linearly dependent on the number of input cascades ($M$). Note that the execution time depends on the value of $\mu$ parameters, which indicates how community-structured the network is. As the value of $\mu$ increases, the community structure in the network fades away, so the cascade will be more likely to survive since it can easily traverse through the network, despite a community. The average length of cascades for datasets with different community structures in \tabref{tab:lfr_chars} indicates this signification. Therefore, as the modular structure of the network decreases, the average length of cascades ($\overline {{n_c}}$) increases and the execution time of DANI increases.

\subsection{Comparison Results}\label{sec:results}

In this section, we provide a complete comparison of the performance of DANI compared to competing methods on synthetic and real datasets in \ref{sec:compsyn} and \ref{sec:compreal}, respectively, and while comparing the performance of these methods, we examine the impact of dataset characteristics on the performance of these models.

\subsubsection{Synthetic datasets}\label{sec:compsyn}

Five synthetic networks with different community structures are shown in \tabref{tab:lfr_chars}. We have generated nine sets of cascades on each of these networks. A total of $45$ different synthetic datasets are used to compare and study the effect of community structure and the number of observed cascades on the performance of different models.

\begin{figure}[!t]
    \centering
    \begin{subfigure}{0.245\textwidth}
        \includegraphics[width=\textwidth]{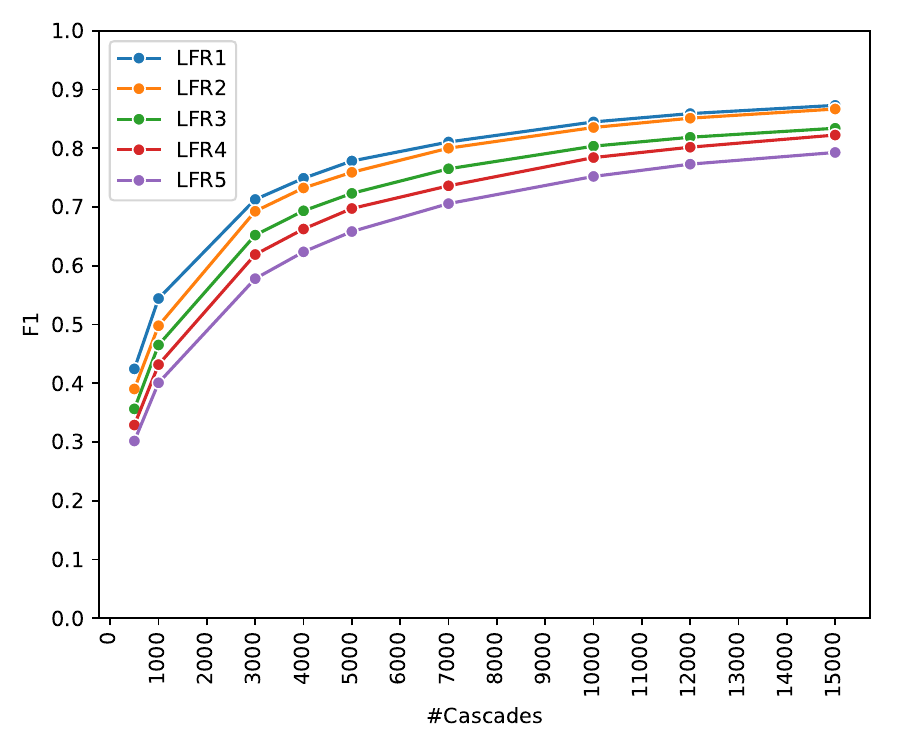}
        \caption{DANI}
        \label{fig:lfr-cas-f1-dani}
    \end{subfigure}
    \hfill
    \begin{subfigure}{0.245\textwidth}
        \includegraphics[width=\textwidth]{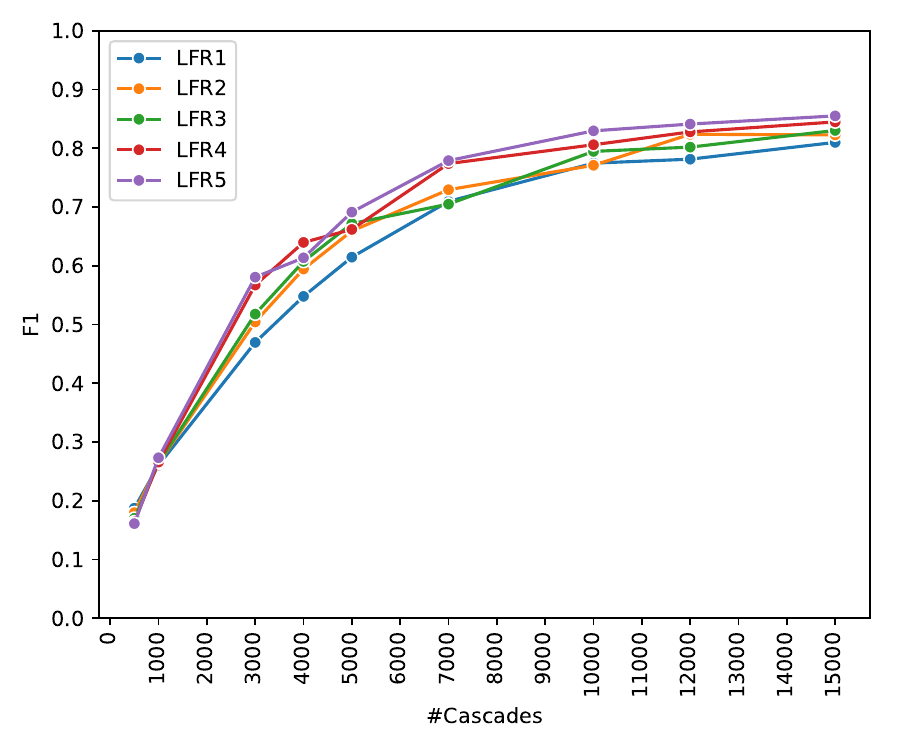}
        \caption{NetInf}
        \label{fig:lfr-cas-f1-NetInf}
    \end{subfigure}
    \hfill
    \begin{subfigure}{0.245\textwidth}
        \includegraphics[width=\textwidth]{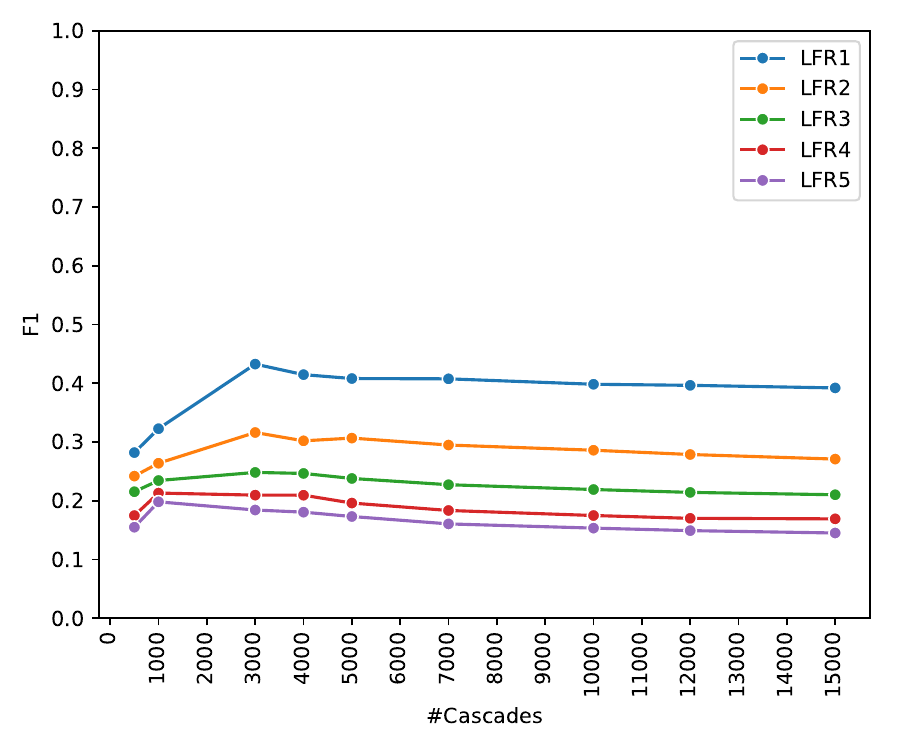}
        \caption{DNE}
        \label{fig:lfr-cas-f1-dne}
    \end{subfigure}
    \hfill
    \begin{subfigure}{0.245\textwidth}
        \includegraphics[width=\textwidth]{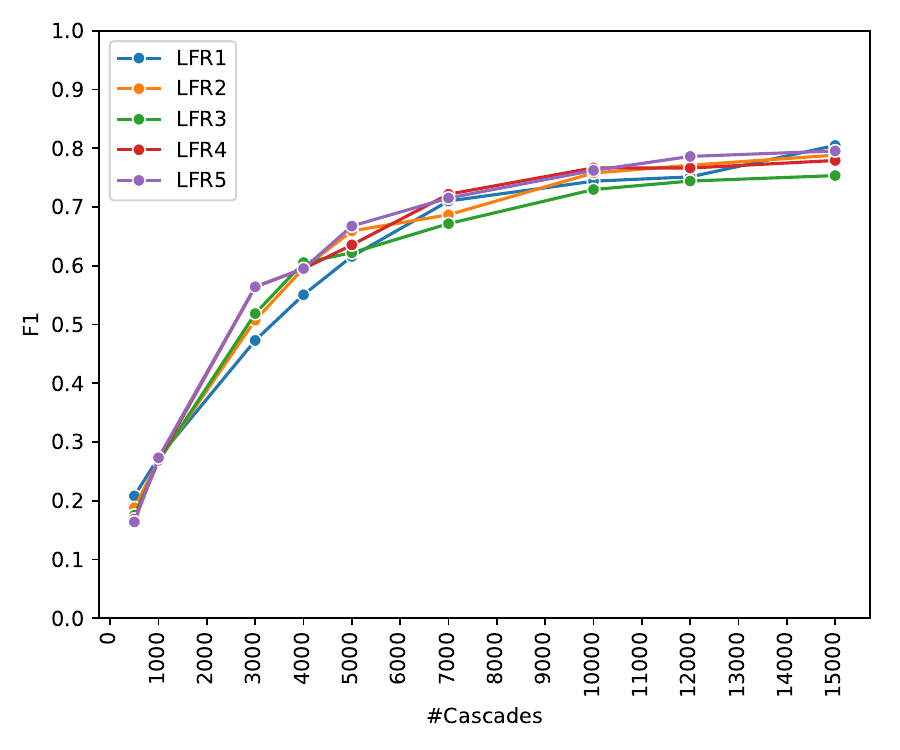}
        \caption{MultiTree}
        \label{fig:lfr-cas-f1-multitree}
    \end{subfigure}
    \hfill
    \begin{subfigure}{0.245\textwidth}
        \includegraphics[width=\textwidth]{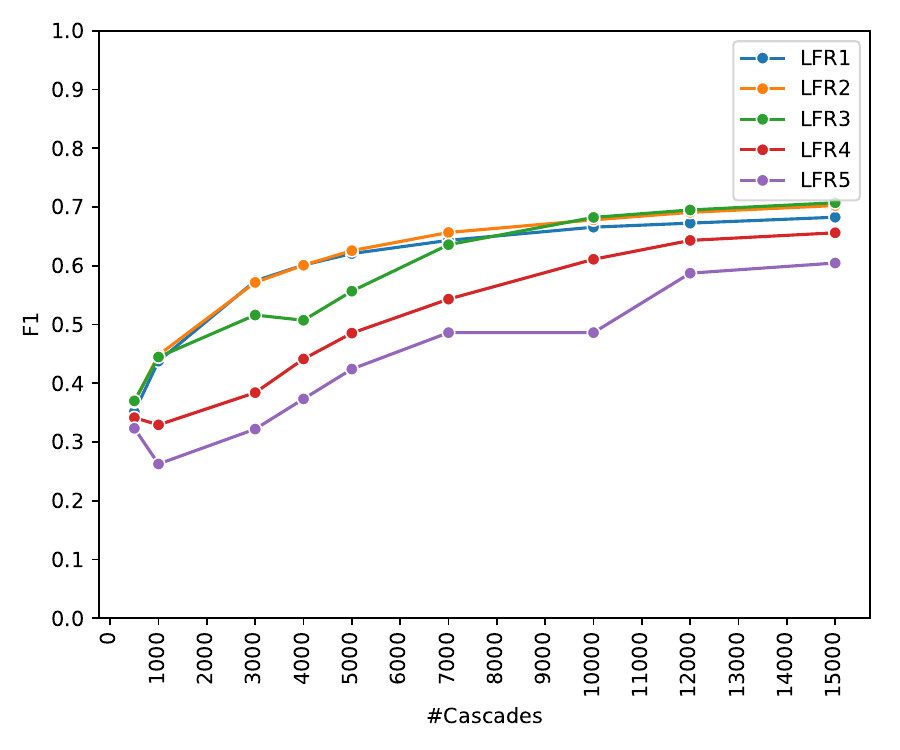}
        \caption{NetRate}
        \label{fig:lfr-cas-f1-netrate}
    \end{subfigure}
    \hfill
    \begin{subfigure}{0.245\textwidth}
        \includegraphics[width=\textwidth]{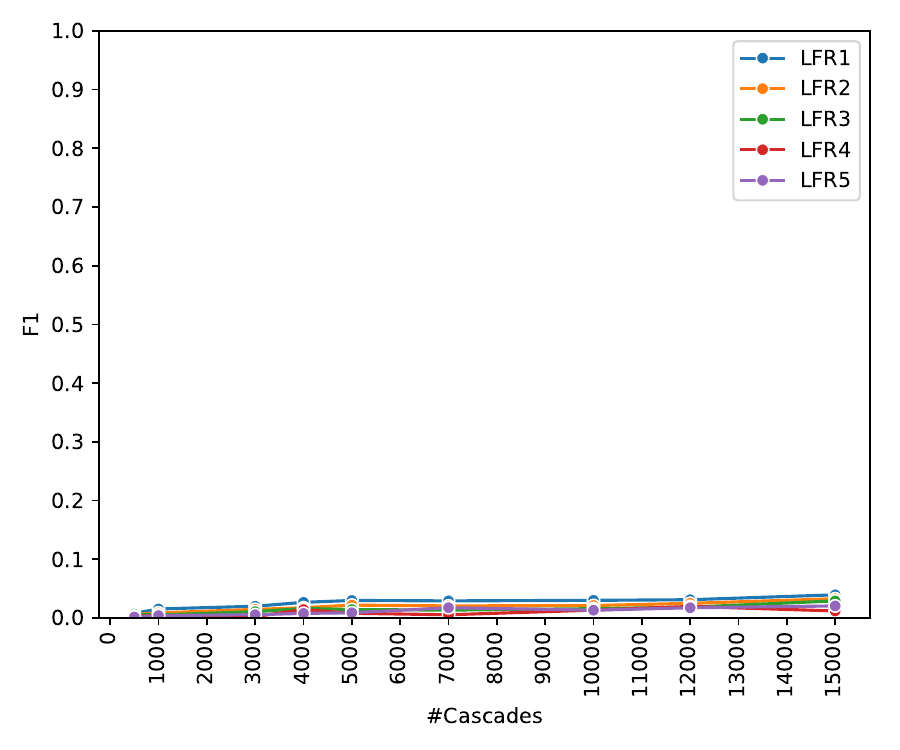}
        \caption{KEBC}
        \label{fig:lfr-cas-f1-kebc}
    \end{subfigure}
    \hfill
    \begin{subfigure}{0.245\textwidth}
        \includegraphics[width=\textwidth]{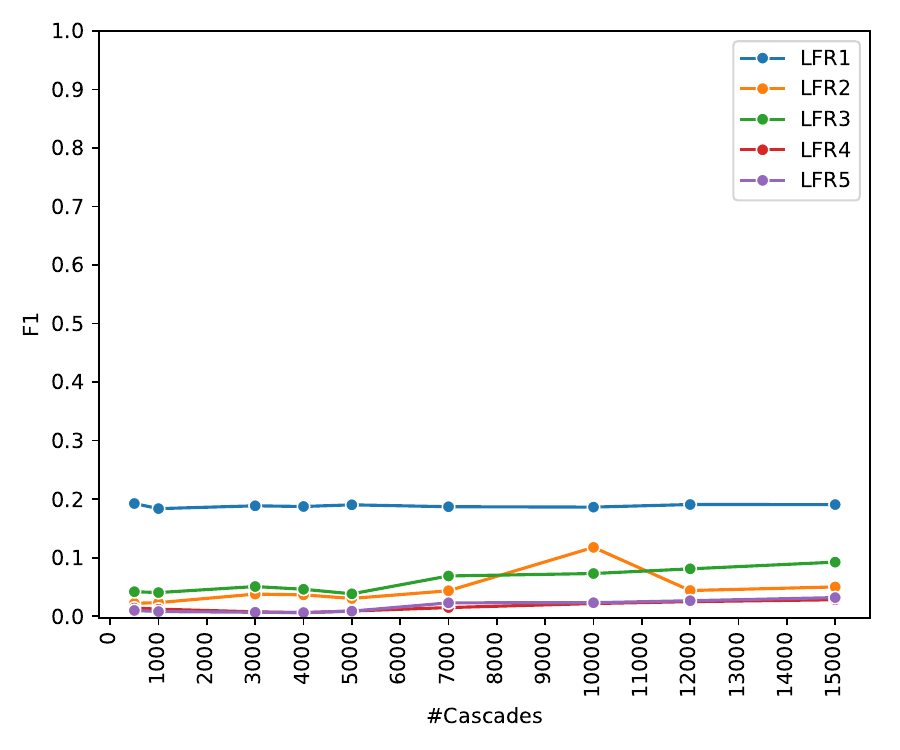}
        \caption{REFINE}
        \label{fig:lfr-cas-f1-refine}
    \end{subfigure}
    \hfill
    \begin{subfigure}{0.245\textwidth}
        \includegraphics[width=\textwidth]{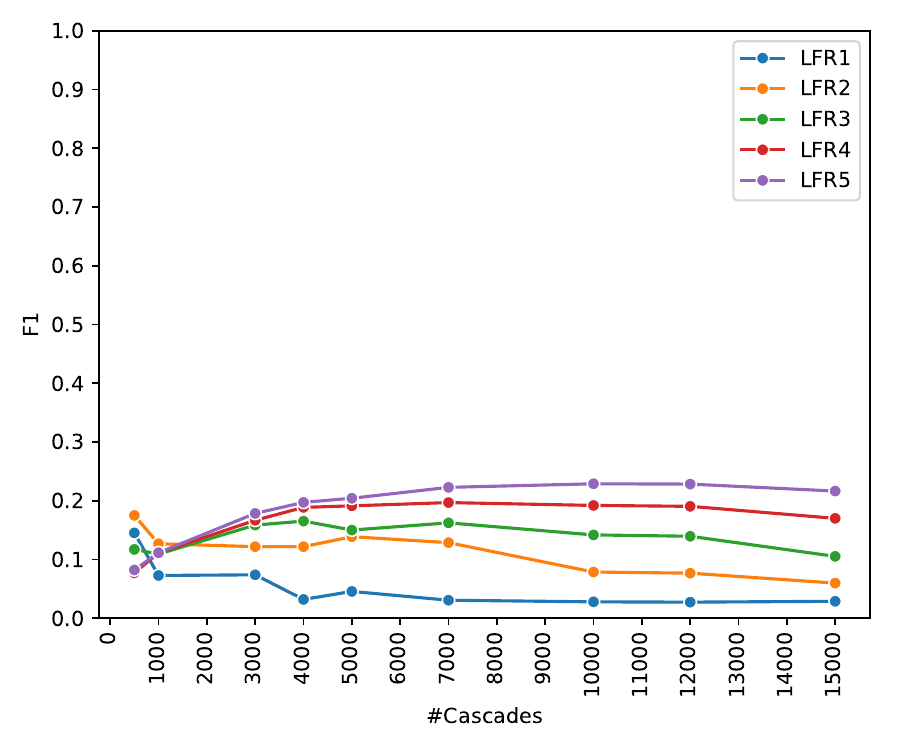}
        \caption{PBI}
        \label{fig:lfr-cas-f1-pbi}
    \end{subfigure}
    \caption{Effect of observation size on F1 score and community structure of LFR Datasets.}
    \label{fig:lfr-cas-f1}
\end{figure}

In \figref{fig:lfr-cas-f1}, we analyzed the effect of the number of observed cascades and community structure on the performance of DANI and other methods for the F1 metric. As illustrated, DANI and NetInf perform better than the other methods. NetInf starts with lower performance than DANI, with fewer observed cascades but performs slightly better when the number of observed cascades increases. DANI and NetRate can perform better than other methods with fewer observed cascades, while DANI performs better in a higher number of cascades. DANI, NetInf, MultiTree, and NetRate performances depend on the number of observed cascades, while the performance of other models seem independent from the number of observed cascades. The performance of some of these models, like DANI, DNE, NetRate, KEBC, and REFINE, directly depends on the value of $\mu$, which indicates how community-structured the network is. Note that the performance of REFINE collapses when the community structure of the underlying network fades away. Since methods like NetInf and MultiTree do not consider structure properties, their performance is independent of $\mu$. The performance of PBI is inversely dependent on the value of $\mu$. As the value of $\mu$ increases, the community structure fades away, and the network becomes more non-modular. PBI performs better on less community-structured networks. KEBC needs prior knowledge of the network and works when there is a limited list of possible candidates for the edges. Since we don't have any prior knowledge, we consider all of the edges of a complete graph as potential candidates, so KEBC cannot perform well in our problem context. It’s worth mentioning that KEBC and PBI consider graphs with undirected edges, while other methods consider directed edges. This causes lower performance for these models.

\begin{figure}[!t]
    \centering
    \begin{subfigure}{0.245\textwidth}
        \includegraphics[width=\textwidth]{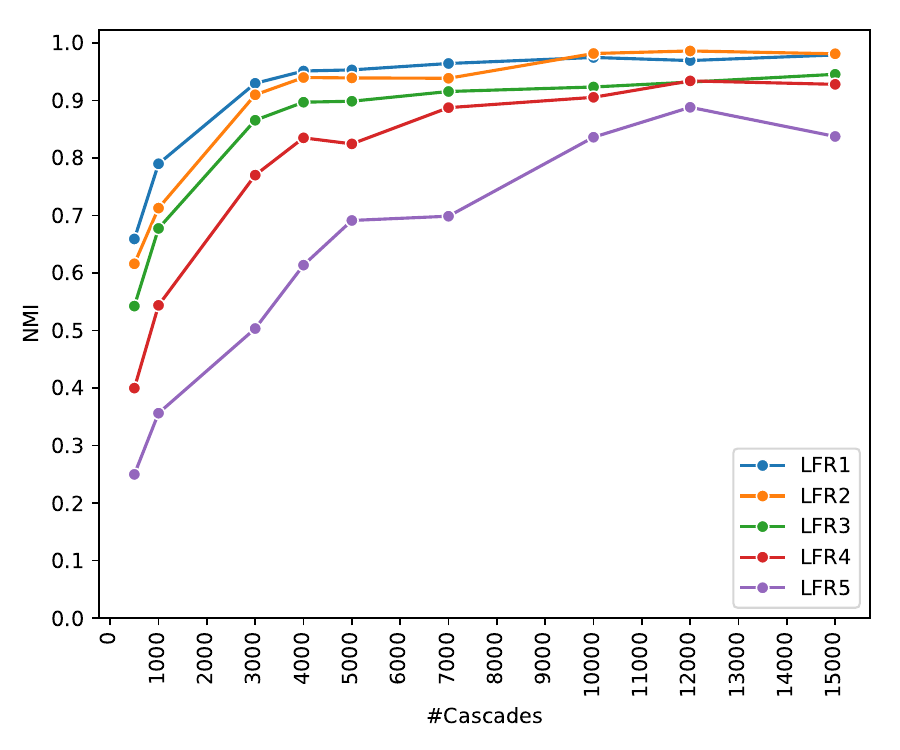}
        \caption{DANI}
        \label{fig:lfr-cas-nmi-dani}
    \end{subfigure}
    \hfill
    \begin{subfigure}{0.245\textwidth}
        \includegraphics[width=\textwidth]{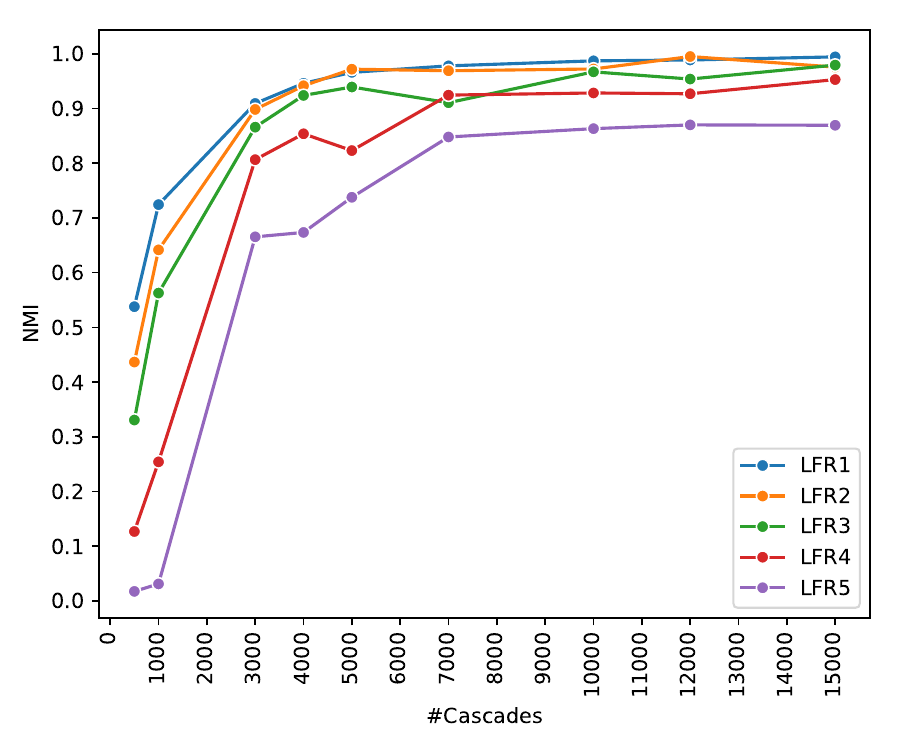}
        \caption{NetInf}
        \label{fig:lfr-cas-nmi-NetInf}
    \end{subfigure}
    \hfill
    \begin{subfigure}{0.245\textwidth}
        \includegraphics[width=\textwidth]{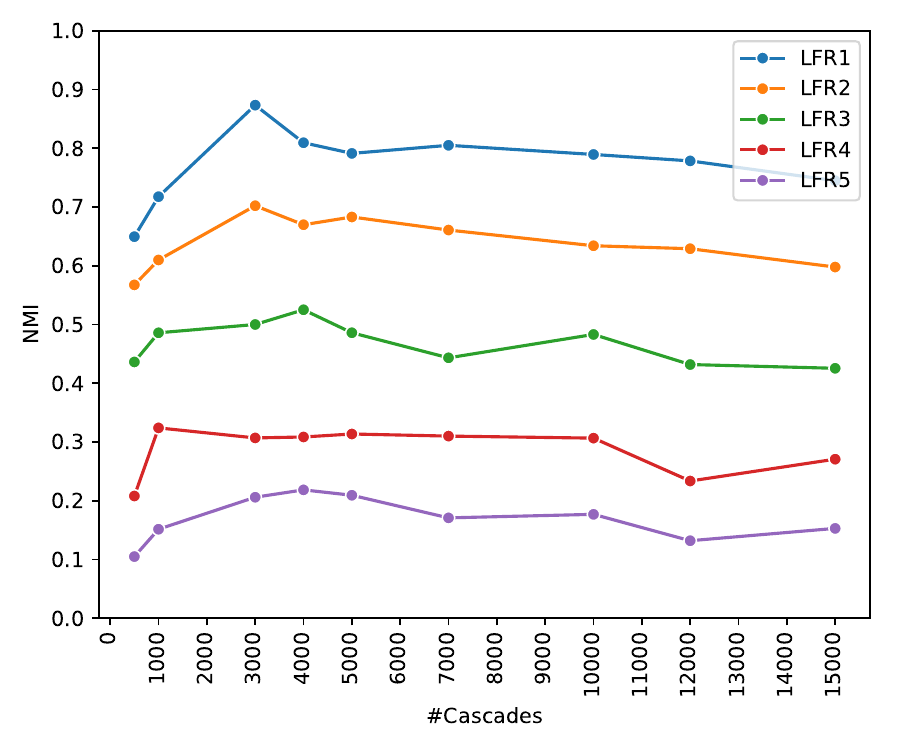}
        \caption{DNE}
        \label{fig:lfr-cas-nmi-dne}
    \end{subfigure}
    \hfill
    \begin{subfigure}{0.245\textwidth}
        \includegraphics[width=\textwidth]{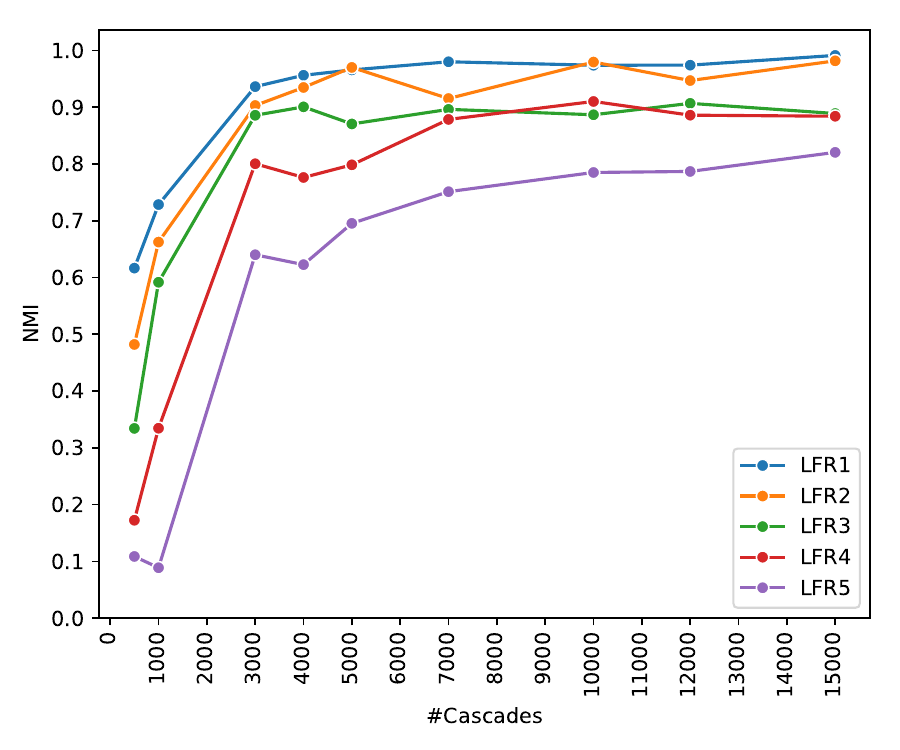}
        \caption{MultiTree}
        \label{fig:lfr-cas-nmi-multitree}
    \end{subfigure}
    \hfill
    \begin{subfigure}{0.245\textwidth}
        \includegraphics[width=\textwidth]{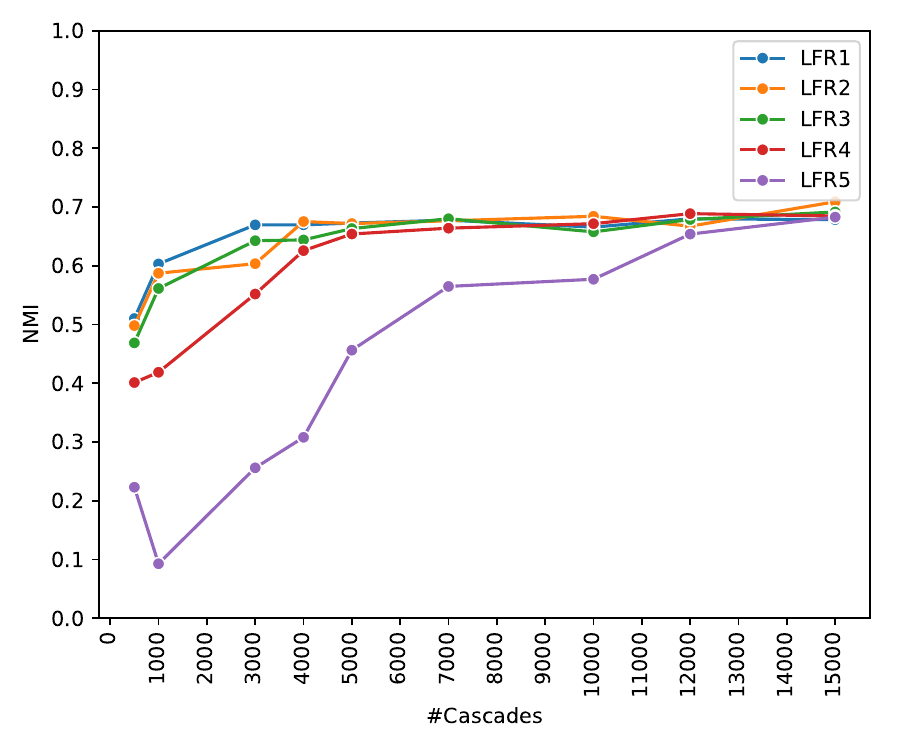}
        \caption{NetRate}
        \label{fig:lfr-cas-nmi-netrate}
    \end{subfigure}
    \hfill
    \begin{subfigure}{0.245\textwidth}
        \includegraphics[width=\textwidth]{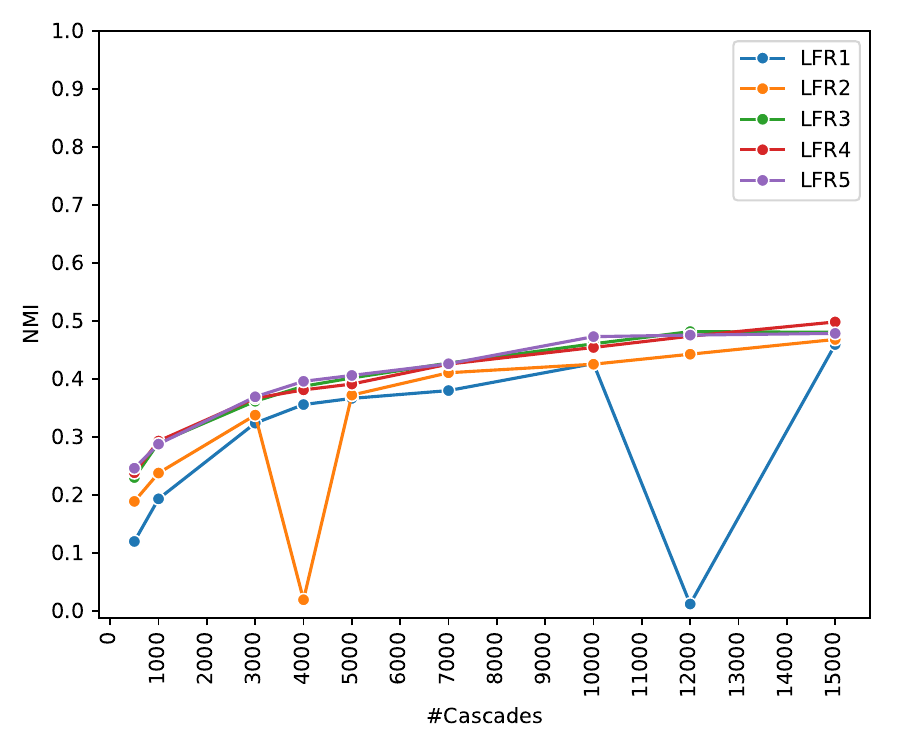}
        \caption{KEBC}
        \label{fig:lfr-cas-nmi-kebc}
    \end{subfigure}
    \hfill
    \begin{subfigure}{0.245\textwidth}
        \includegraphics[width=\textwidth]{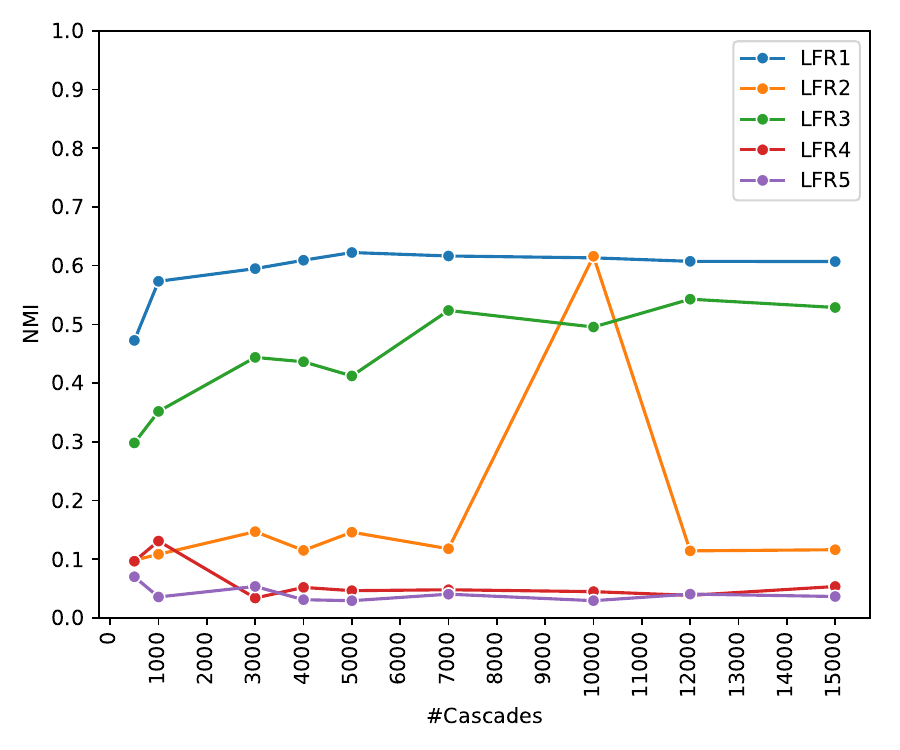}
        \caption{REFINE}
        \label{fig:lfr-cas-nmi-refine}
    \end{subfigure}
    \hfill
    \begin{subfigure}{0.245\textwidth}
        \includegraphics[width=\textwidth]{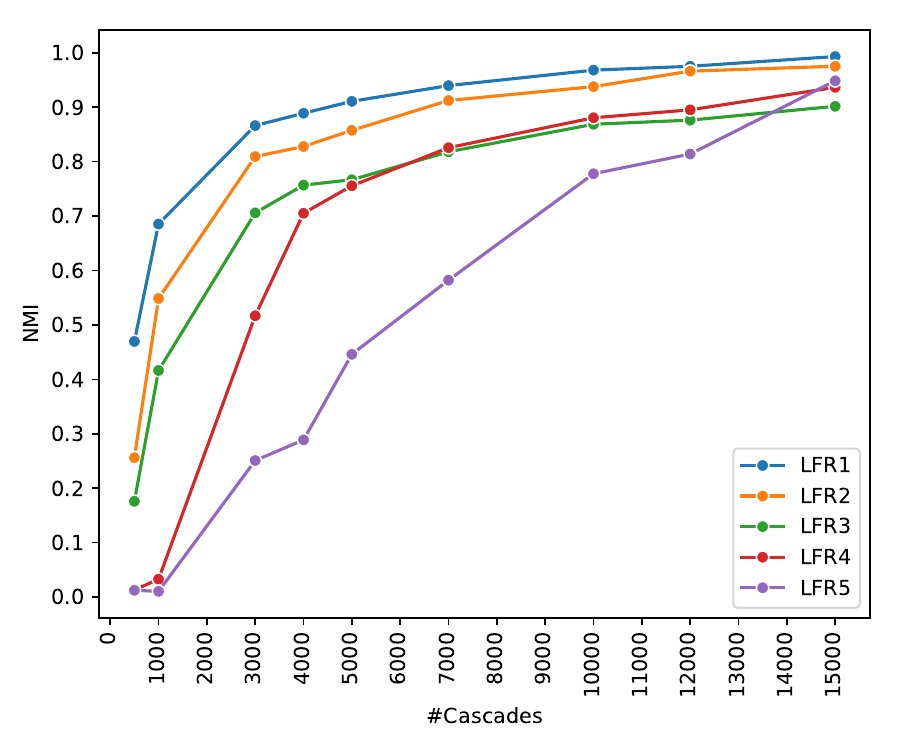}
        \caption{PBI}
        \label{fig:lfr-cas-nmi-pbi}
    \end{subfigure}
    \caption{Effect of observation size and community structure on NMI of LFR Datasets.}
    \label{fig:lfr-cas-nmi}
\end{figure}

Similar to what we did for comparing F1, \figref{fig:lfr-cas-nmi} compares NMI with respect to the number of observed cascades and community structure of the networks. Note that we used a third-party method, OSLOM, for community detection, so despite the low inference performance of methods like KEBC, REFINE, and PBI, their NMI is acceptable because NMI is independent of the inferred graph but dependent on the communities detected by OSLOM over the inferred graph. OSLOM can detect communities with a limited number of input links, even on noisy data. Although PBI performs poorly in link inference, it has a very high NMI value. KEBC and REFINE also have similar conditions. These NMI results on models with poor link inference performance are fake because of the post-processing done by OSLOM. So, we limit our study to the methods that perform well in link inference with respect to F1. As shown in \figref{fig:lfr-cas-nmi-dani}, \figref{fig:lfr-cas-nmi-NetInf}, \figref{fig:lfr-cas-nmi-dne}, \figref{fig:lfr-cas-nmi-multitree}, and \figref{fig:lfr-cas-nmi-netrate}, when the network is more community-structured, the value of NMI increases. The value of NMI depends on F1 because OSLOM can detect communities better when the inferred network is more accurate, so NMI has a similar trend to F1 in \figref{fig:lfr-cas-f1}.

\begin{figure}[ht]
    \centering
    \begin{subfigure}{0.245\textwidth}
        \includegraphics[width=\textwidth]{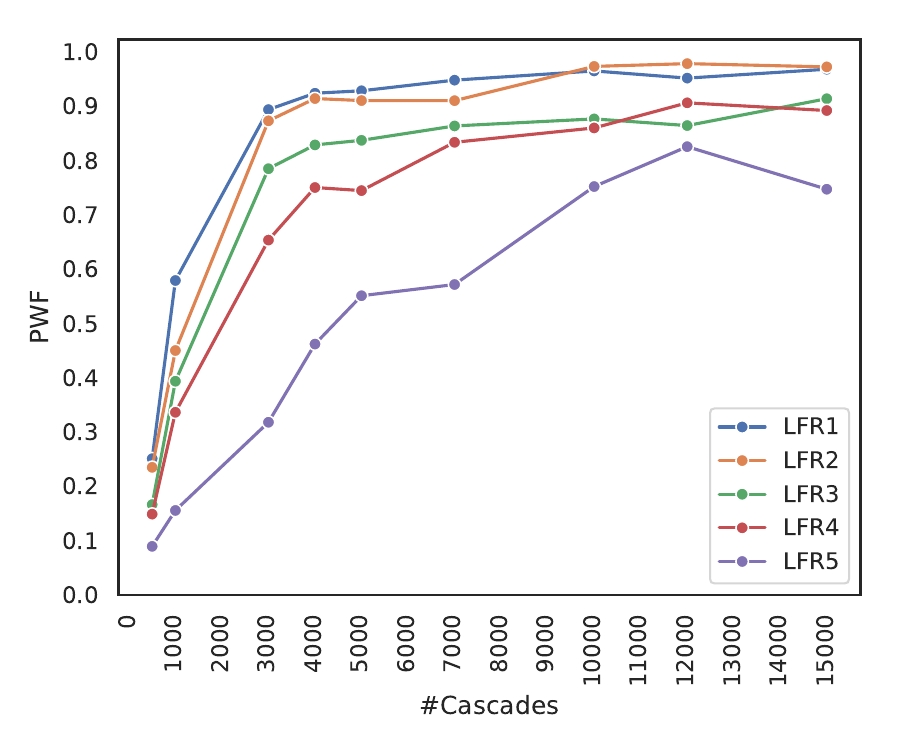}
        \caption{DANI}
        \label{fig:lfr-cas-pwf-dani}
    \end{subfigure}
    \hfill
    \begin{subfigure}{0.245\textwidth}
        \includegraphics[width=\textwidth]{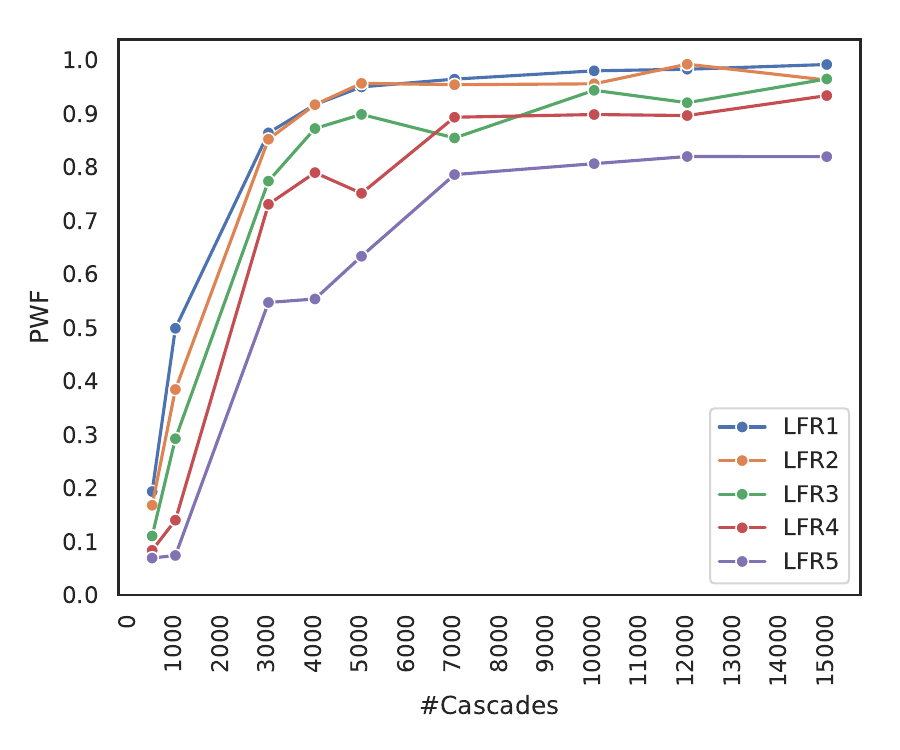}
        \caption{NetInf}
        \label{fig:lfr-cas-pwf-NetInf}
    \end{subfigure}
    \hfill
    \begin{subfigure}{0.245\textwidth}
        \includegraphics[width=\textwidth]{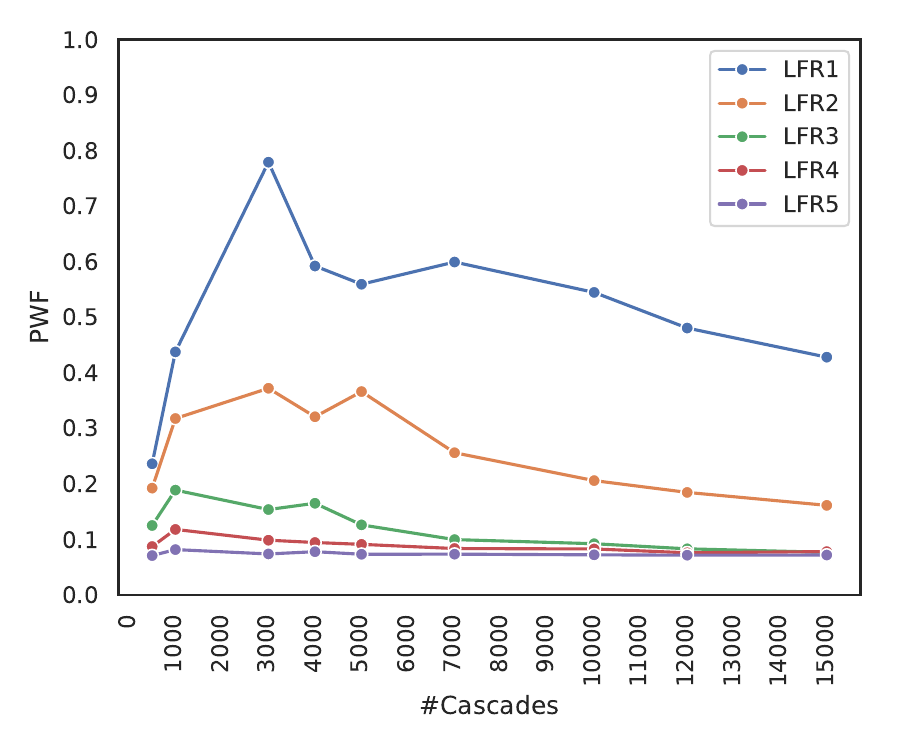}
        \caption{DNE}
        \label{fig:lfr-cas-pwf-dne}
    \end{subfigure}
    \hfill
    \begin{subfigure}{0.245\textwidth}
        \includegraphics[width=\textwidth]{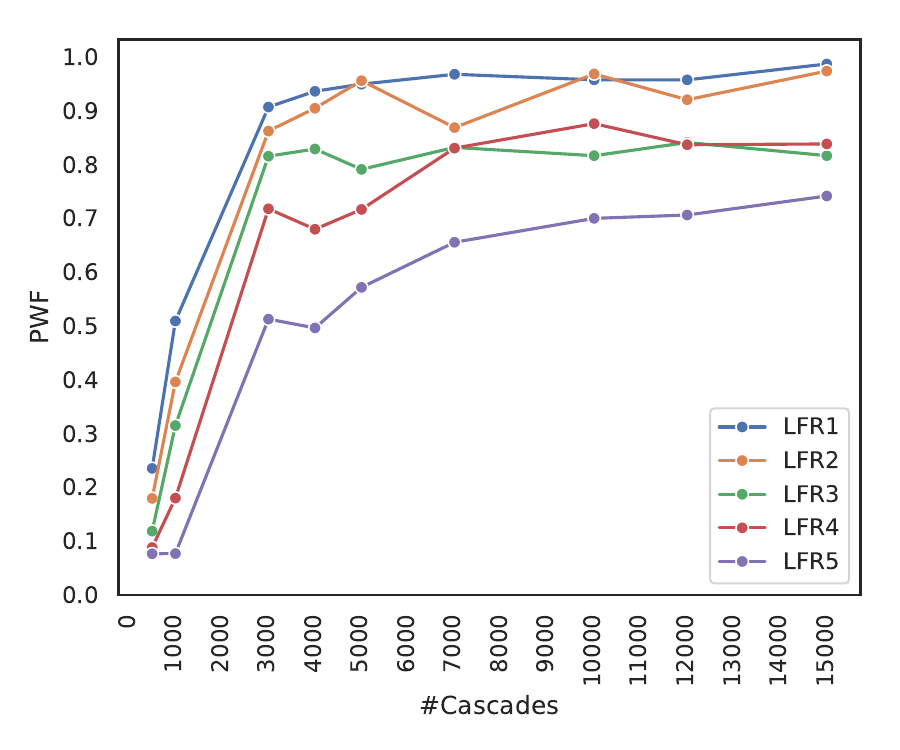}
        \caption{MultiTree}
        \label{fig:lfr-cas-pwf-multitree}
    \end{subfigure}
    \hfill
    \begin{subfigure}{0.245\textwidth}
        \includegraphics[width=\textwidth]{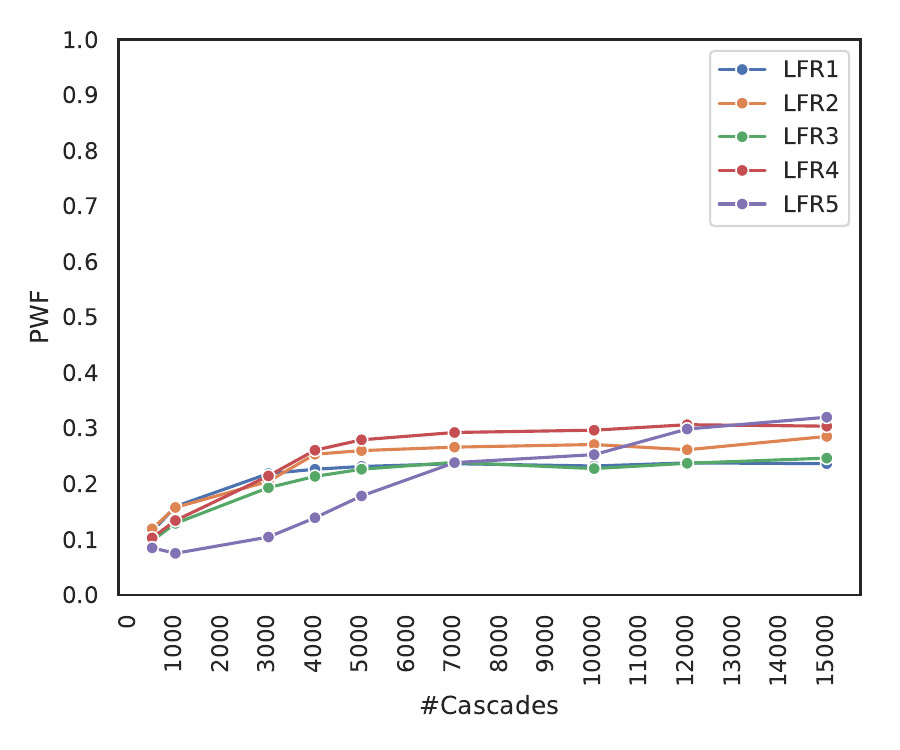}
        \caption{NetRate}
        \label{fig:lfr-cas-pwf-netrate}
    \end{subfigure}
    \hfill
    \begin{subfigure}{0.245\textwidth}
        \includegraphics[width=\textwidth]{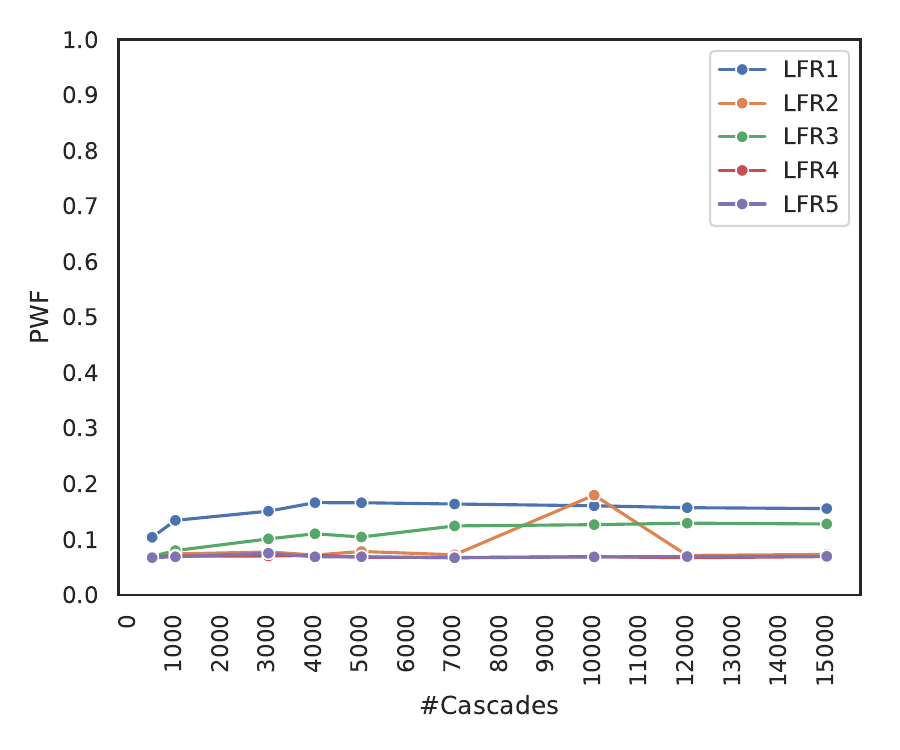}
        \caption{KEBC}
        \label{fig:lfr-cas-pwf-kebc}
    \end{subfigure}
    \hfill
    \begin{subfigure}{0.245\textwidth}
        \includegraphics[width=\textwidth]{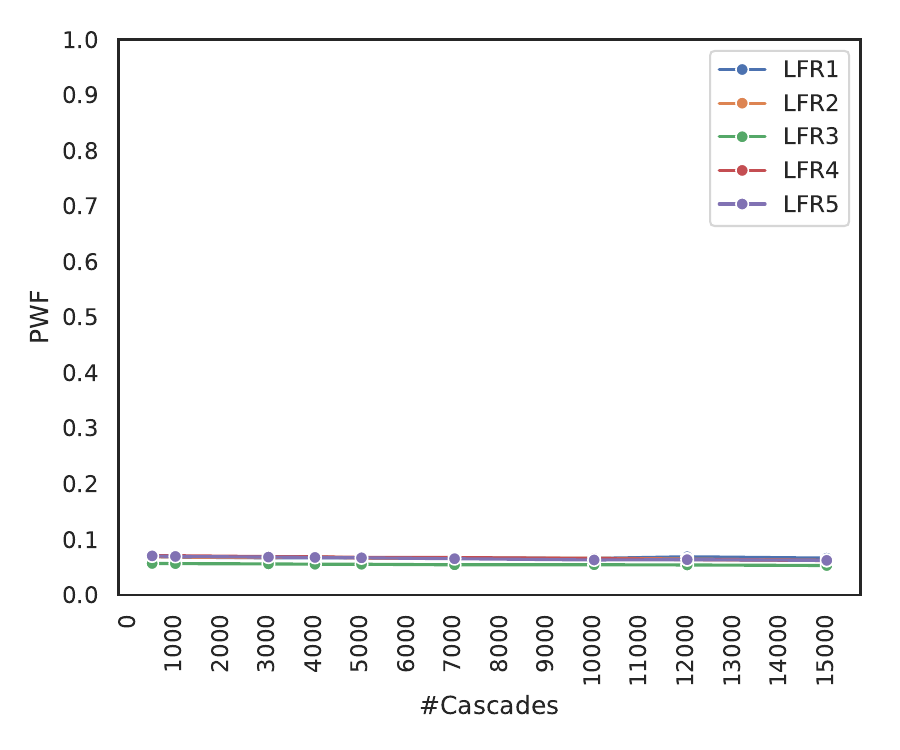}
        \caption{REFINE}
        \label{fig:lfr-cas-pwf-refine}
    \end{subfigure}
    \hfill
    \begin{subfigure}{0.245\textwidth}
        \includegraphics[width=\textwidth]{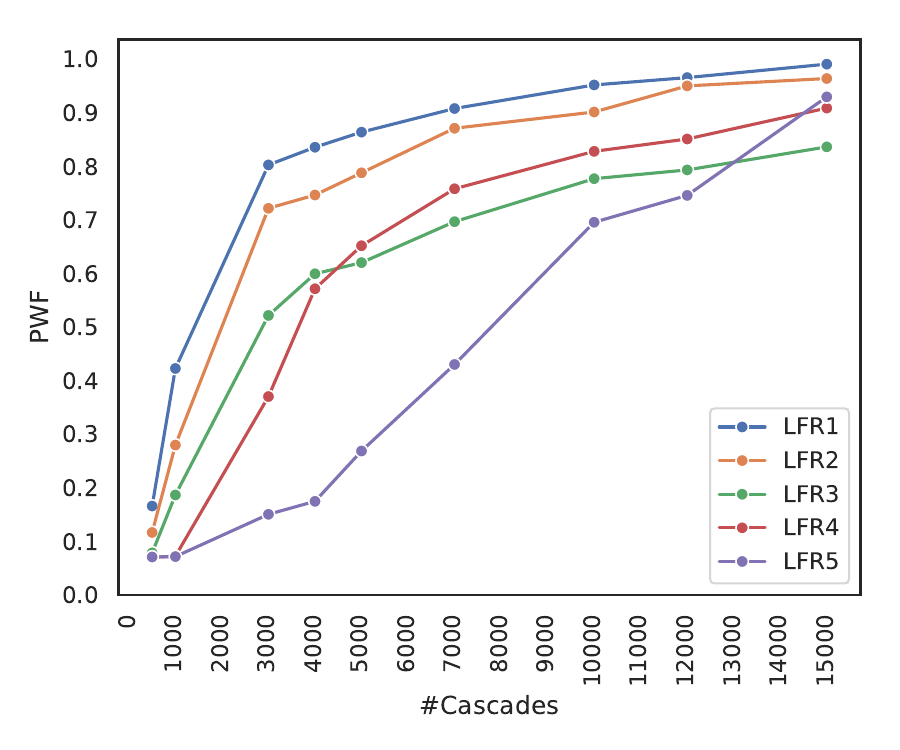}
        \caption{PBI}
        \label{fig:lfr-cas-pwf-pbi}
    \end{subfigure}
    \caption{Effect of observation size and community structure on PWF of LFR Datasets.}
    \label{fig:lfr-cas-pwf}
\end{figure}

In \figref{fig:lfr-cas-pwf}, we compare models with respect to PWF. Similar to what was said about NMI, PWF is a measure that does not depend directly on the inferred network but is calculated from the communities detected by OSLOM. So, the good results of PBI with respect to PWF are fake too. In other methods except for NetRate, PWF increases when the network is more community-structured. In NetRate, initially with a limited number of observed cascades, PWF is better for more structured networks, but as the number of observed cascades increases, networks with less community structure become superior with respect to PWF. Cascades in networks with less community structure are longer on average, as shown in \tabref{tab:lfr_chars}, so a more robust survival analysis can be done, and the probability of correct community detection increases. As shown, the performance of other models is proportional to their performance in link inference. DANI, MultiTree, and NetInf perform better, while DANI can perform well with a small number of observations despite the two others.

Previously, we investigated the effect of the number of observed cascades on the performance of the models. We saw that with the increase in the number of cascades, metrics for most of the models converge to a specific value, and increasing it more than that does not cause a significant change in the performance of the models. From now on, in comparing the performance of models on synthetic datasets, we focus on the datasets with observation sets having $10000$ cascades.

\begin{figure}[!t]
    \centering
    \begin{subfigure}{0.32\textwidth}
        \includegraphics[width=\textwidth]{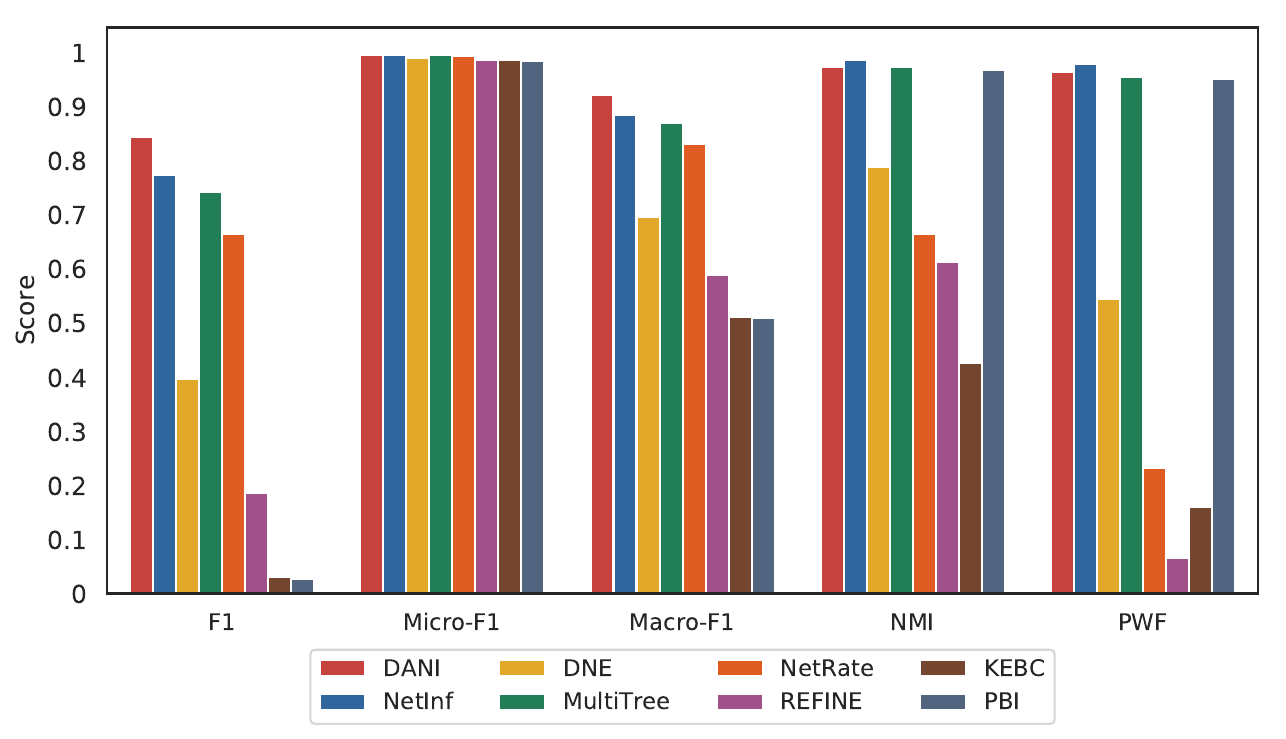}
        \caption{LFR1}
        \label{fig:lfr-performance-1}
    \end{subfigure}
    \hfill
    \begin{subfigure}{0.32\textwidth}
        \includegraphics[width=\textwidth]{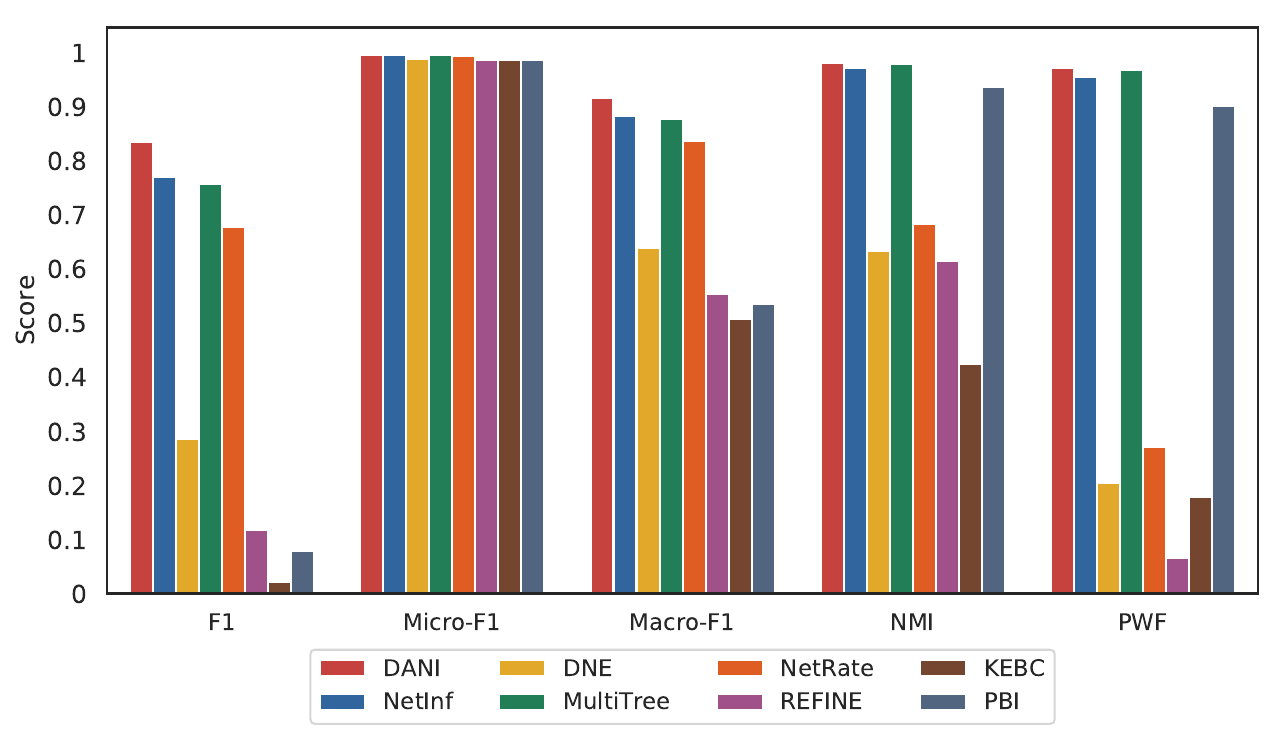}
        \caption{LFR2}
        \label{fig:lfr-performance-2}
    \end{subfigure}
    \hfill
    \begin{subfigure}{0.32\textwidth}
        \includegraphics[width=\textwidth]{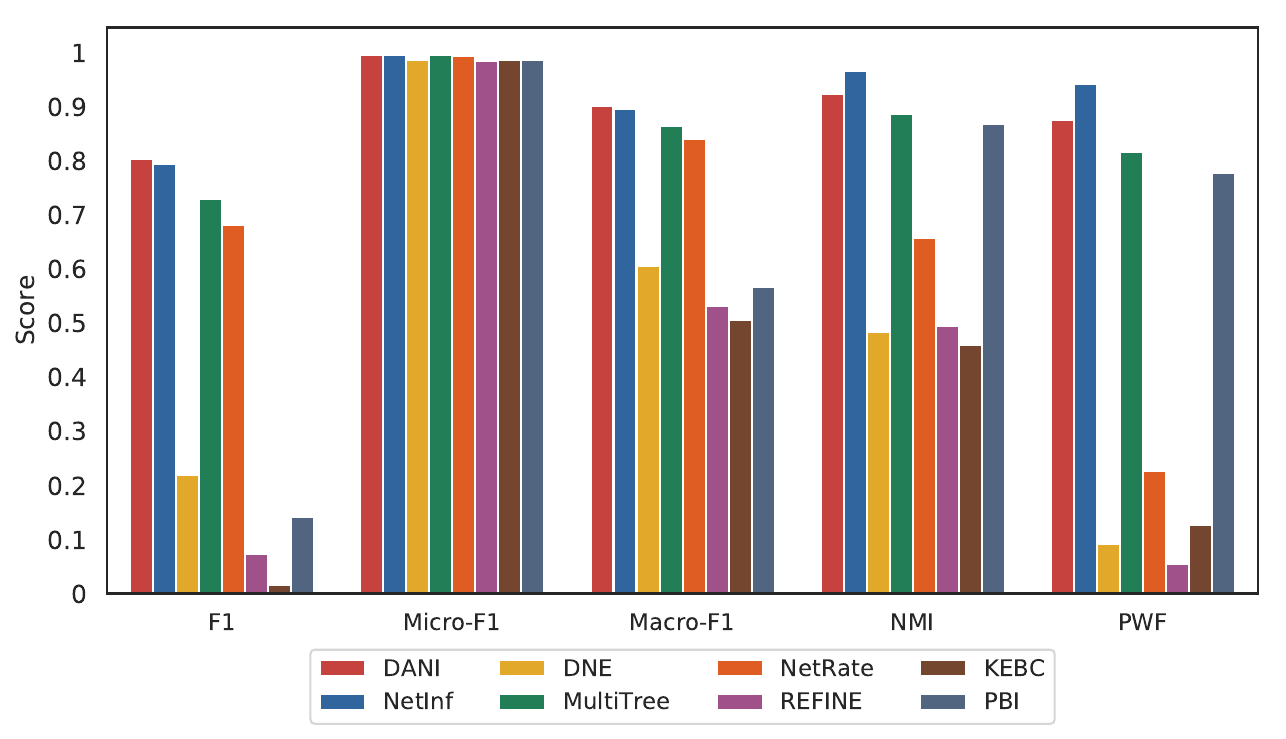}
        \caption{LFR3}
        \label{fig:lfr-performance-3}
    \end{subfigure}
    \hfill
    \begin{subfigure}{0.32\textwidth}
        \includegraphics[width=\textwidth]{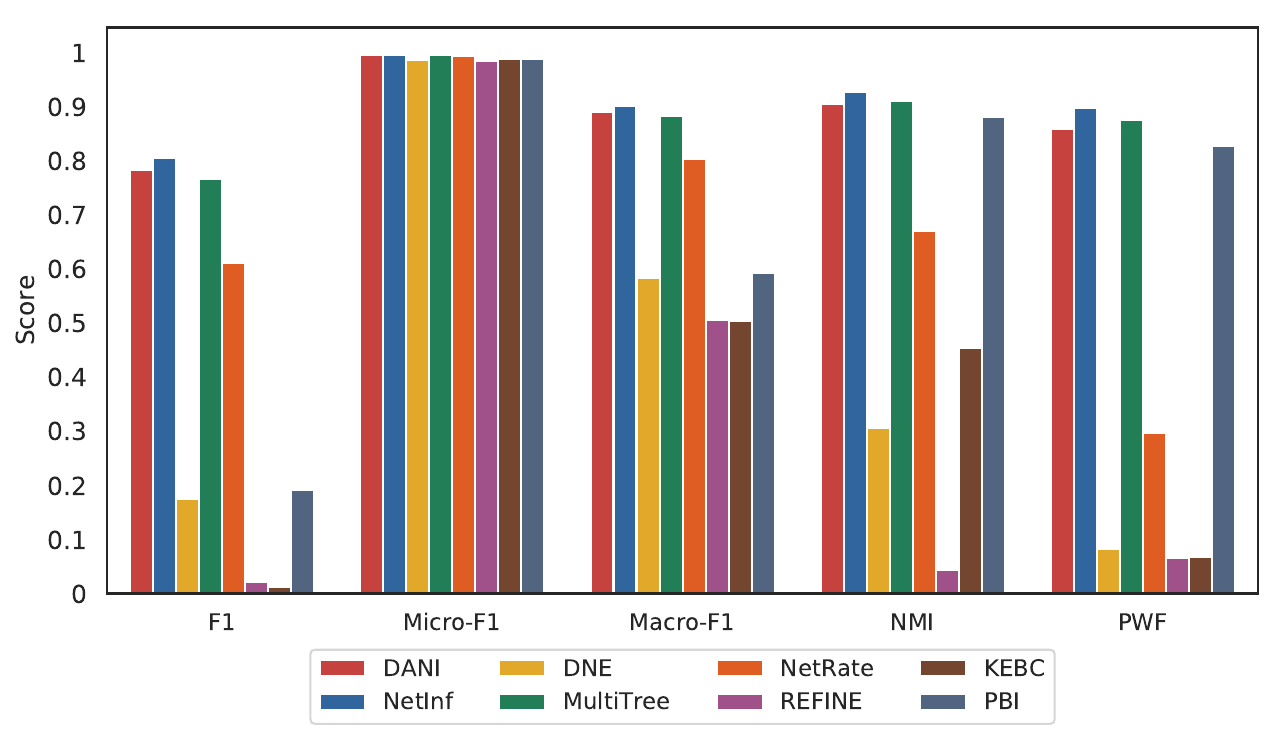}
        \caption{LFR4}
        \label{fig:lfr-performance-4}
    \end{subfigure}
    \begin{subfigure}{0.32\textwidth}
        \includegraphics[width=\textwidth]{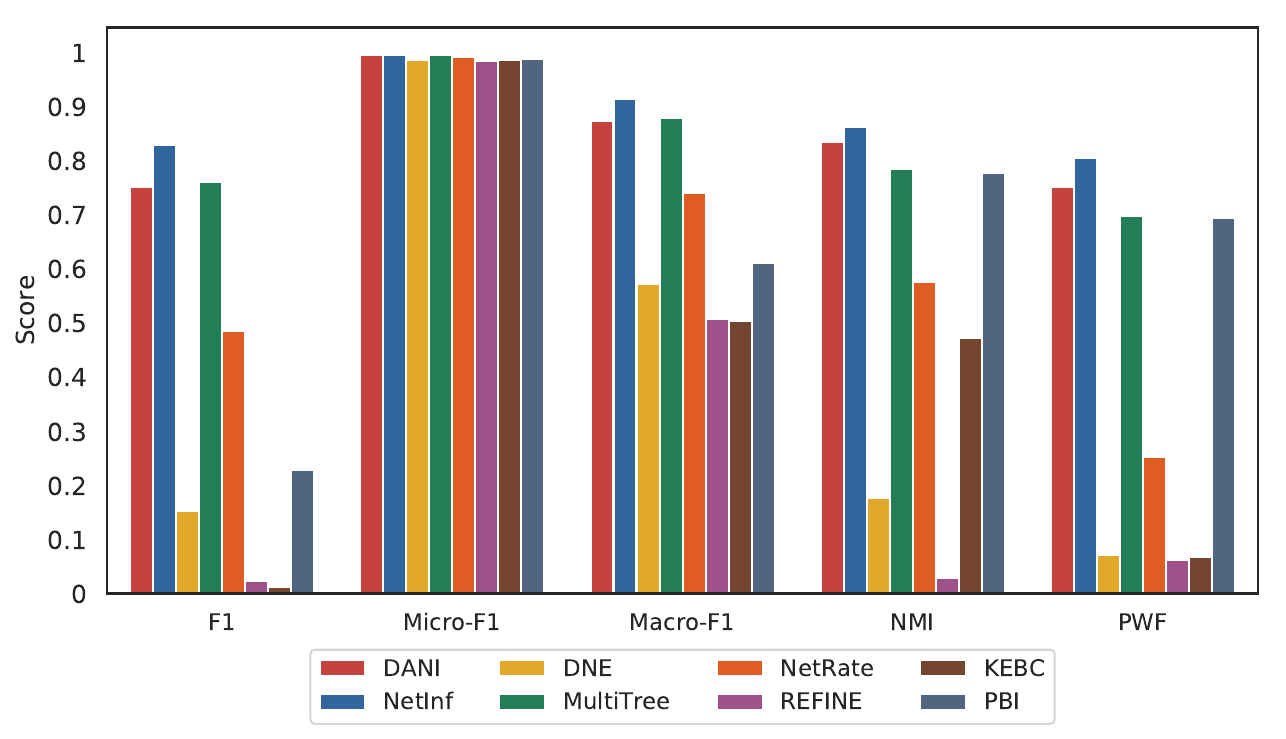}
        \caption{LFR5}
        \label{fig:lfr-performance-5}
    \end{subfigure}
    \caption{Metric scores on five different LFR networks with $10000$ cascades.}
    \label{performance_lfr}
\end{figure}

In \figref{performance_lfr} and \tabref{tab:lfrresult}, we have conducted a comprehensive study on the performance of the models on the synthetic datasets by observing $10000$ cascades. The metric value of the best model in each dataset is bolded. In the task of link inference with respect to metrics defined in \ref{subsec:metrics_1}, DANI performs better than the other in LFR1, LFR2, and LFR3 in which the networks are more modular, while NetInf performs better in LFR4 and LFR5, in which network is getting close to random network and modular structure fades out since NetInf is not sensitive to structural topology of underlying network. As discussed before, we used the Jensen-Shannon distance to compare the difference in the distribution of nodes' degrees in the ground truth and inferred network. In all datasets, DANI has the smallest distance. In other criteria, NMI, PWF, $\Delta$ACS, $\Delta$Cnd, $\Delta$CN, $\Delta\rho$, and $\Delta$CC, In most of them, DANI is one of the top models.  Note that in this category of metrics, methods with poor link inference performance are not acceptable since OSLOM abilities may cause them.

\begin{table}[!b]
\centering
\footnotesize
\caption{Result of DANI and other related works on LFR datasets}
\label{tab:lfrresult}
\begin{tabularx}{\textwidth}{Ycccccccccccc}
\hline\hline
                  & \textbf{Method} & \textbf{F1}   & \textbf{MF1}  & \textbf{mF1}   & \textbf{JS}   & \textbf{NMI}  & \textbf{PWF}  & \textbf{$\boldsymbol{\Delta}$ACS} & \textbf{$\boldsymbol{\Delta}$Cnd} & \textbf{$\boldsymbol{\Delta}$CN} & \textbf{$\boldsymbol{\Delta\rho}$} & \textbf{$\boldsymbol{\Delta}$CC}  \\ 
\hline\hline
\multirow{8}{*}{LFR1} & DANI            & \textbf{0.84} & \textbf{0.92} & \textbf{0.998} & \textbf{0.27} & 0.97          & 0.96          & \textbf{0.00}                     & 0.43                              & \textbf{0.00}                    & \textbf{0.01}                      & 0.32                              \\ 
\cline{2-13}
                  & NetInf          & 0.77          & 0.89          & 0.997          & 0.38          & \textbf{0.99} & \textbf{0.98} & \textbf{0.00}                     & 0.80                              & \textbf{0.00}                    & 0.07                               & 0.33                              \\ 
\cline{2-13}
                  & DNE             & 0.40          & 0.70          & 0.991          & 0.62          & 0.79          & 0.54          & 0.17                              & 4.71                              & 0.15                             & 0.23                               & 0.64                              \\ 
\cline{2-13}
                  & MultiTree       & 0.74          & 0.87          & 0.996          & 0.4           & 0.97          & 0.96          & \textbf{0.00}                     & 1.00                              & \textbf{0.00}                    & 0.09                               & 0.27                              \\ 
\cline{2-13}
                  & NetRate         & 0.67          & 0.83          & 0.995          & 0.48          & 0.67          & 0.23          & \textbf{0.00}                     & 1.44                              & 0.32                             & 0.43                               & \textbf{0.04}                     \\ 
\cline{2-13}
                  & REFINE          & 0.19          & 0.59          & 0.987          & 0.66          & 0.61          & 0.07          & 0.03                              & \textbf{0.35}                     & 0.40                             & 2.08                               & 2.15                              \\ 
\cline{2-13}
                  & KEBC            & 0.03          & 0.51          & 0.988          & 0.72          & 0.43          & 0.16          & 0.93                              & 15.99                             & 0.58                             & 1.00                               & 0.95                              \\ 
\cline{2-13}
                  & PBI             & 0.03          & 0.51          & 0.986          & 0.37          & 0.97          & 0.95          & \textbf{0.00}                     & 1.07                              & \textbf{0.00}                    & 0.25                               & 0.33                              \\ 
\hline
\multirow{8}{*}{LFR2} & DANI            & \textbf{0.84} & \textbf{0.92} & \textbf{0.982} & \textbf{0.24} & 0.98          & \textbf{0.97} & 0.03                              & \textbf{0.28}                     & \textbf{0.00}                    & 0.10                               & 0.30                              \\ 
\cline{2-13}
                  & NetInf          & 0.77          & 0.88          & 0.972          & 0.39          & 0.97          & 0.96          & \textbf{0.00}                     & 0.60                              & \textbf{0.00}                    & 0.15                               & 0.10                              \\ 
\cline{2-13}
                  & DNE             & 0.29          & 0.64          & 0.634          & 0.67          & 0.63          & 0.21          & 0.26                              & 2.73                              & 0.34                             & \textbf{0.03}                      & 0.99                              \\ 
\cline{2-13}
                  & MultiTree       & 0.76          & 0.88          & 0.980          & 0.40          & \textbf{0.98} & 0.97          & \textbf{0.00}                     & 0.57                              & \textbf{0.00}                    & 0.15                               & 0.10                              \\ 
\cline{2-13}
                  & NetRate         & 0.68          & 0.84          & 0.684          & 0.48          & 0.68          & 0.27          & \textbf{0.00}                     & 0.83                              & 0.29                             & 0.47                               & \textbf{0.06}                     \\ 
\cline{2-13}
                  & REFINE          & 0.12          & 0.56          & 0.616          & 0.64          & 0.62          & 0.07          & \textbf{0.00}                     & 0.46                              & 0.37                             & 1.77                               & 3.00                              \\ 
\cline{2-13}
                  & KEBC            & 0.02          & 0.51          & 0.425          & 0.70          & 0.43          & 0.18          & 0.93                              & 7.14                              & 0.58                             & 1.00                               & 0.96                              \\ 
\cline{2-13}
                  & PBI             & 0.08          & 0.54          & 0.938          & 0.39          & 0.94          & 0.90          & 0.03                              & 0.54                              & \textbf{0.00}                    & 0.24                               & 0.41                              \\ 
\hline
\multirow{8}{*}{LFR3} & DANI            & \textbf{0.80} & \textbf{0.90} & \textbf{0.997} & \textbf{0.26} & 0.92          & 0.88          & 0.03                              & 0.25                              & \textbf{0.00}                    & \textbf{0.03}                      & 0.24                              \\ 
\cline{2-13}
                  & NetInf          & 0.79          & 0.90          & 0.997          & 0.36          & \textbf{0.97} & \textbf{0.94} & \textbf{0.00}                     & 0.33                              & \textbf{0.00}                    & 0.15                               & \textbf{0.04}                     \\ 
\cline{2-13}
                  & DNE             & 0.22          & 0.61          & 0.988          & 0.70          & 0.48          & 0.09          & 0.33                              & 1.65                              & 0.46                             & 0.08                               & 1.21                              \\ 
\cline{2-13}
                  & MultiTree       & 0.73          & 0.86          & 0.996          & 0.41          & 0.89          & 0.82          & 0.03                              & 0.47                              & 0.01                             & 0.25                               & 0.06                              \\ 
\cline{2-13}
                  & NetRate         & 0.68          & 0.84          & 0.995          & 0.51          & 0.66          & 0.23          & 0.03                              & 0.79                              & 0.28                             & 0.39                               & 0.15                              \\ 
\cline{2-13}
                  & REFINE          & 0.07          & 0.53          & 0.985          & 0.62          & 0.50          & 0.05          & 0.03                              & 0.24                              & 0.30                             & 1.39                               & 3.30                              \\ 
\cline{2-13}
                  & KEBC            & 0.02          & 0.51          & 0.988          & 0.68          & 0.46          & 0.13          & 0.92                              & 4.28                              & 0.54                             & 1.00                               & 0.94                              \\ 
\cline{2-13}
                  & PBI             & 0.14          & 0.57          & 0.987          & 0.41          & 0.87          & 0.78          & \textbf{0.00}                     & \textbf{0.24}                     & \textbf{0.00}                    & 0.31                               & 0.46                              \\ 
\hline
\multirow{8}{*}{LFR4} & DANI            & 0.78          & 0.89          & 0.997          & \textbf{0.27} & 0.91          & 0.86          & 0.03                              & 0.29                              & \textbf{0.00}                    & 0.03                               & 0.33                              \\ 
\cline{2-13}
                  & NetInf          & \textbf{0.81} & \textbf{0.90} & \textbf{0.997} & 0.35          & \textbf{0.93} & \textbf{0.90} & \textbf{0.00}                     & 0.31                              & \textbf{0.00}                    & 0.17                               & \textbf{0.00}                     \\ 
\cline{2-13}
                  & DNE             & 0.17          & 0.58          & 0.987          & 0.73          & 0.31          & 0.08          & 0.09                              & 1.38                              & 0.57                             & \textbf{0.01}                      & 3.79                              \\ 
\cline{2-13}
                  & MultiTree       & 0.77          & 0.88          & 0.996          & 0.38          & 0.91          & 0.88          & \textbf{0.00}                     & 0.33                              & 0.01                             & 0.21                               & 0.02                              \\ 
\cline{2-13}
                  & NetRate         & 0.61          & 0.80          & 0.994          & 0.53          & 0.67          & 0.30          & \textbf{0.00}                     & 0.60                              & 0.26                             & 0.30                               & 0.22                              \\ 
\cline{2-13}
                  & REFINE          & 0.02          & 0.51          & 0.985          & 0.75          & 0.04          & 0.07          & 4.80                              & 0.46                              & 0.42                             & 0.16                               & 7.72                              \\ 
\cline{2-13}
                  & KEBC            & 0.01          & 0.50          & 0.988          & 0.70          & 0.45          & 0.07          & 0.94                              & 2.86                              & 0.53                             & 1.00                               & 0.93                              \\ 
\cline{2-13}
                  & PBI             & 0.19          & 0.59          & 0.988          & 0.39          & 0.88          & 0.83          & \textbf{0.00}                     & \textbf{0.25}                     & \textbf{0.00}                    & 0.28                               & 0.49                              \\ 
\hline
\multirow{8}{*}{LFR5} & DANI            & 0.75          & 0.88          & 0.996          & \textbf{0.27} & 0.84          & 0.75          & 0.03                              & 0.27                              & \textbf{0.00}                    & \textbf{0.06}                      & 0.57                              \\ 
\cline{2-13}
                  & NetInf          & \textbf{0.83} & \textbf{0.91} & \textbf{0.997} & 0.34          & \textbf{0.86} & \textbf{0.81} & \textbf{0.00}                     & 0.20                              & \textbf{0.00}                    & 0.21                               & 0.12                              \\ 
\cline{2-13}
                  & DNE             & 0.15          & 0.57          & 0.987          & 0.75          & 0.18          & 0.07          & 0.19                              & 0.67                              & 0.62                             & 0.23                               & 6.65                              \\ 
\cline{2-13}
                  & MultiTree       & 0.76          & 0.88          & 0.996          & 0.40          & 0.79          & 0.70          & \textbf{0.00}                     & 0.29                              & 0.01                             & 0.21                               & \textbf{0.10}                     \\ 
\cline{2-13}
                  & NetRate         & 0.49          & 0.74          & 0.992          & 0.48          & 0.58          & 0.25          & 0.07                              & 0.63                              & 0.25                             & 0.07                               & 0.31                              \\ 
\cline{2-13}
                  & REFINE          & 0.02          & 0.51          & 0.985          & 0.73          & 0.03          & 0.06          & 6.25                              & \textbf{0.09}                     & 0.50                             & 0.66                               & 16.72                             \\ 
\cline{2-13}
                  & KEBC            & 0.01          & 0.50          & 0.988          & 0.70          & 0.47          & 0.07          & 0.94                              & 1.88                              & 0.48                             & 1.00                               & 0.87                              \\ 
\cline{2-13}
                  & PBI             & 0.23          & 0.61          & 0.989          & 0.35          & 0.78          & 0.70          & \textbf{0.00}                     & 0.18                              & \textbf{0.00}                    & 0.32                               & 0.47                              \\
\hline\hline
\end{tabularx}
\end{table}

\begin{figure}[t]
    \centering
    \begin{subfigure}{0.245\textwidth}
        \includegraphics[width=\textwidth]{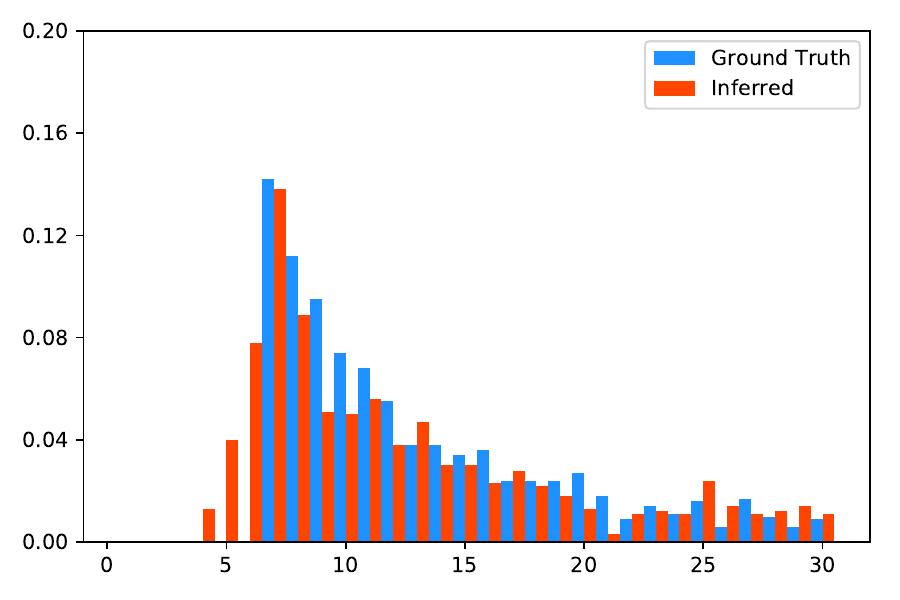}
        \caption{DANI}
        \label{fig:lfr-deg-dani}
    \end{subfigure}
    \hfill
    \begin{subfigure}{0.245\textwidth}
        \includegraphics[width=\textwidth]{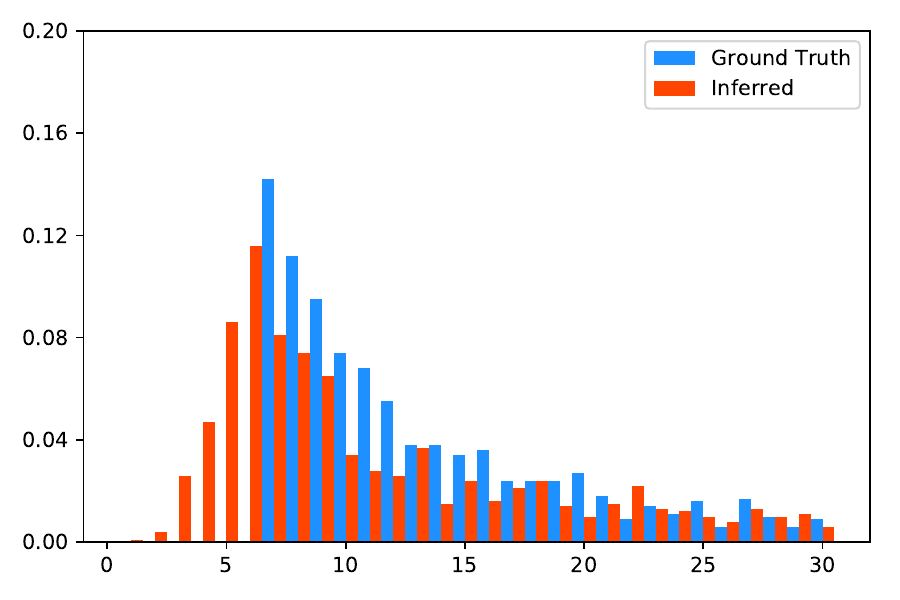}
        \caption{NetInf}
        \label{fig:lfr-deg-NetInf}
    \end{subfigure}
    \hfill
    \begin{subfigure}{0.245\textwidth}
        \includegraphics[width=\textwidth]{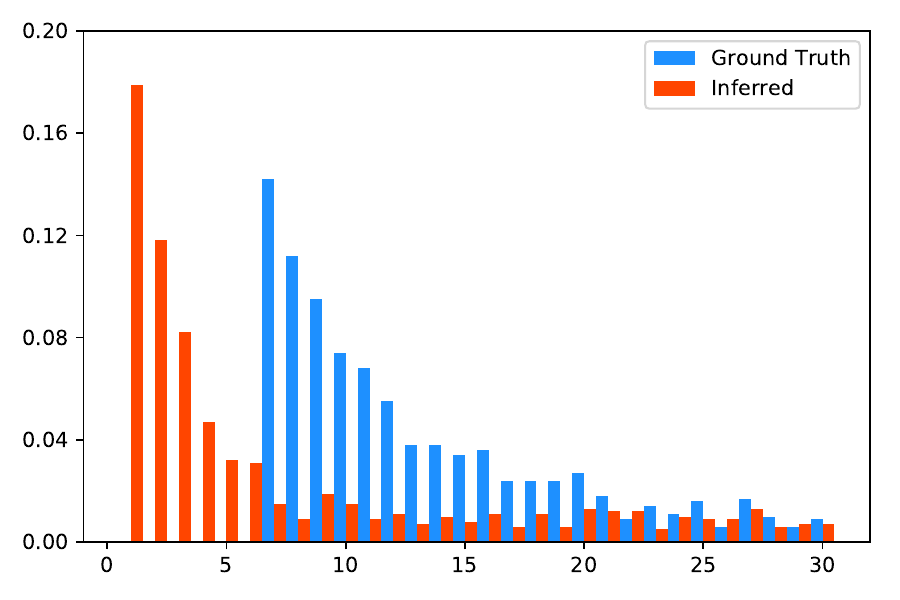}
        \caption{DNE}
        \label{fig:lfr-deg-dne}
    \end{subfigure}
    \hfill
    \begin{subfigure}{0.245\textwidth}
        \includegraphics[width=\textwidth]{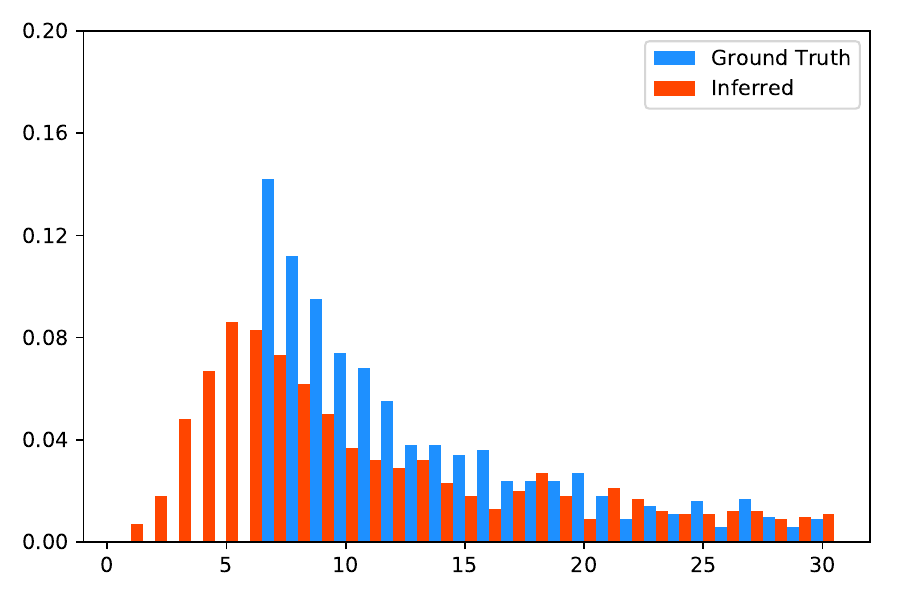}
        \caption{MultiTree}
        \label{fig:lfr-deg-multitree}
    \end{subfigure}
    \hfill
    \begin{subfigure}{0.245\textwidth}
        \includegraphics[width=\textwidth]{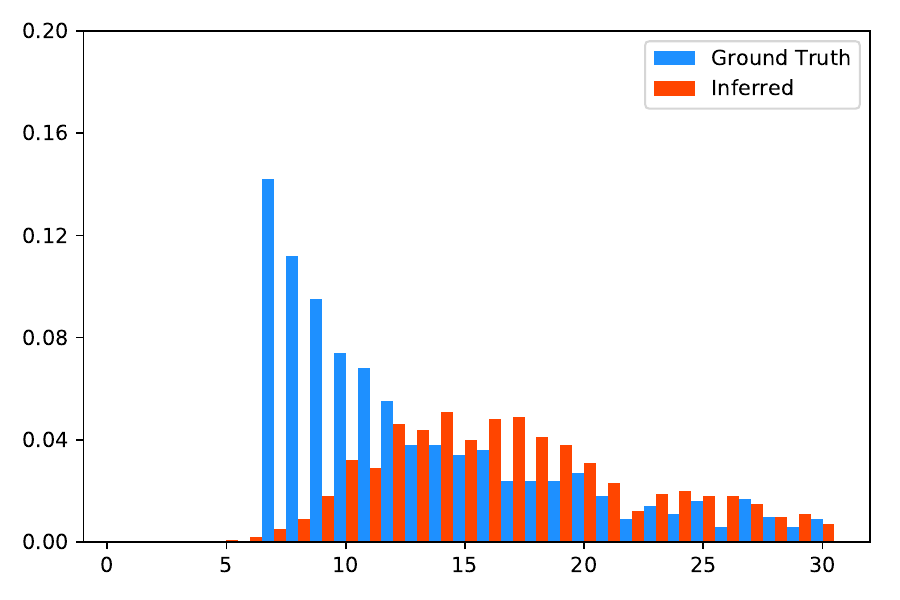}
        \caption{NetRate}
        \label{fig:lfr-deg-netrate}
    \end{subfigure}
    \hfill
    \begin{subfigure}{0.245\textwidth}
        \includegraphics[width=\textwidth]{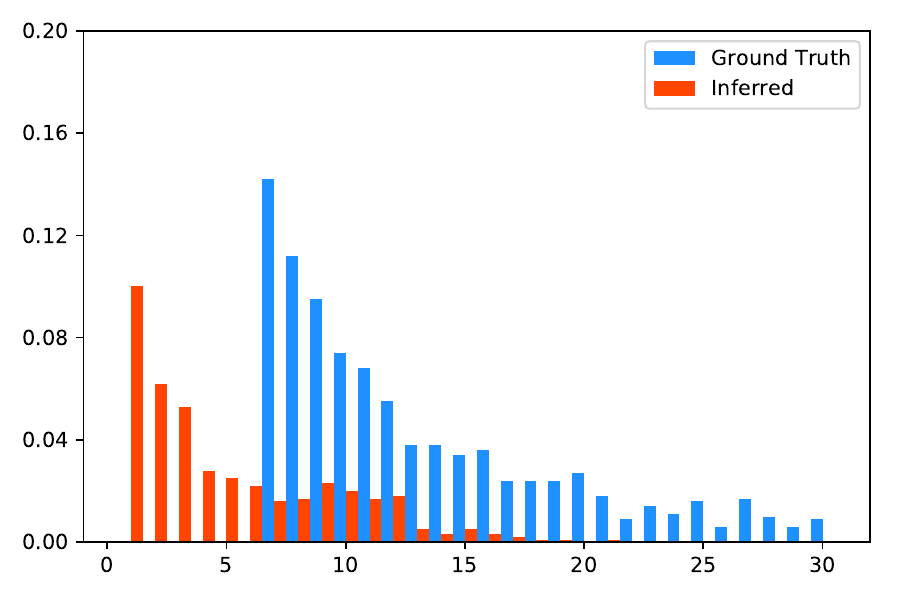}
        \caption{KEBC}
        \label{fig:lfr-deg-kebc}
    \end{subfigure}
    \hfill
    \begin{subfigure}{0.245\textwidth}
        \includegraphics[width=\textwidth]{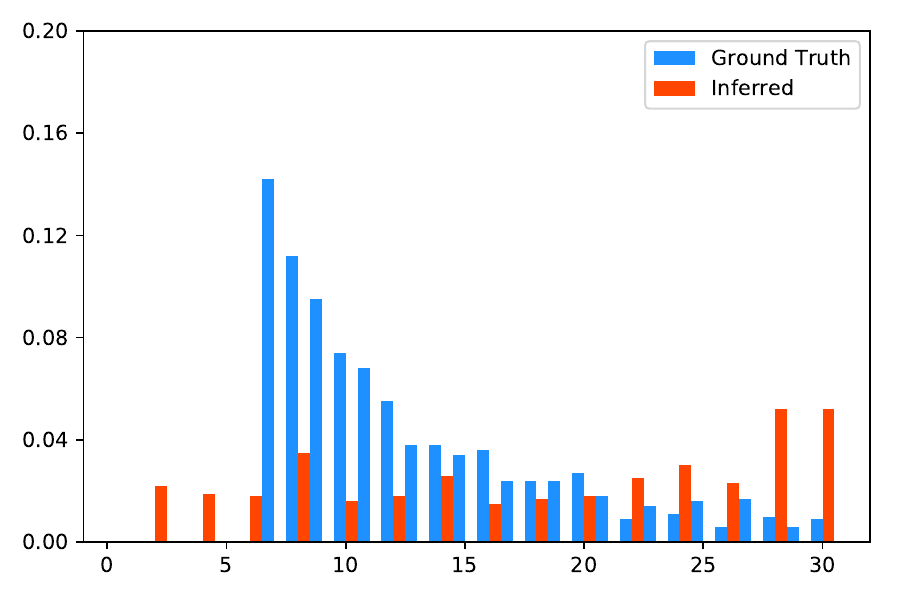}
        \caption{REFINE}
        \label{fig:lfr-deg-refine}
    \end{subfigure}
    \hfill
    \begin{subfigure}{0.245\textwidth}
        \includegraphics[width=\textwidth]{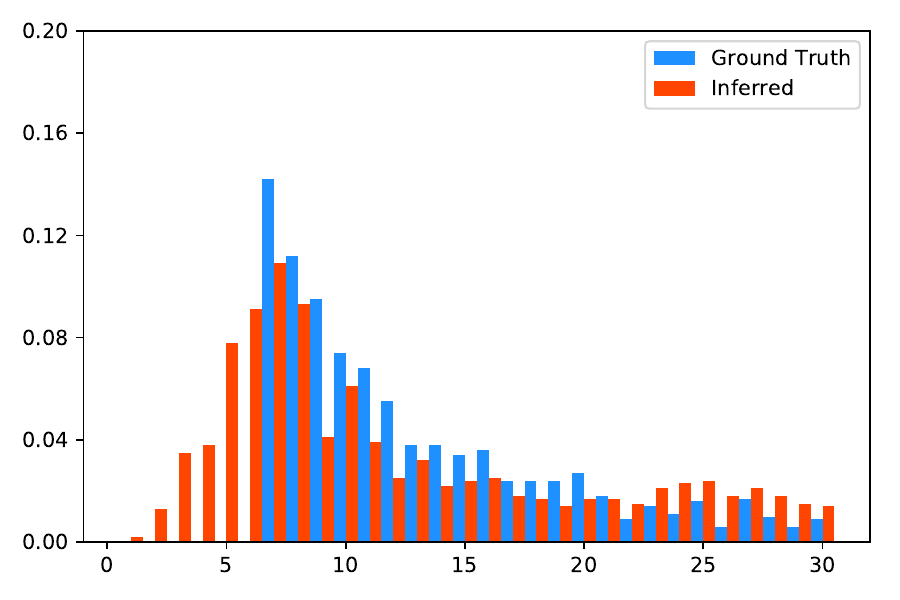}
        \caption{PBI}
        \label{fig:lfr-deg-pbi}
    \end{subfigure}
    \caption{Distribution of nodes degree in ground truth and inferred network of LFR $\mu=0.1$ dataset with 10000 observed cascades.}
    \label{fig:lfr-deg}
\end{figure}

We compare the distributions of nodes' degrees of ground truth and inferred network. The distributions of node degrees can indicate the similarity between the original and predicted networks. We used the Jensen-Shannon distance in \tabref{tab:lfrresult} to measure the similarity of the two distributions. In \figref{fig:lfr-deg}, the statistical distribution of node degrees for LFR1 with 10000 observed cascades is drawn. As illustrated, the distribution of nodes degrees of the network inferred by DANI is the most similar to the ground truth. The distribution of methods like DNE and KEBC tends to zero, while methods like NetInf, MultiTree, NetRate, and PBI tend to have more smooth and normalized distributions. Since REFINE performs poorly in link inference, its distribution is not meaningful.

\subsubsection{Real datasets}\label{sec:compreal}

We evaluate the models on real datasets listed in \tabref{tab:realresult}. Note that NetRate, PBI, and KEBC could not terminate on these datasets in 72 hours, their results on real datasets are not reported. Since REFINE requires allocating $N \times N$ matrices and Twitter has the largest number of nodes ($N=124k$) among other datasets, we don't have REFINE results on Twitter. \tabref{tab:realresult} displays the behavior of the methods on real datasets.

LinkedIn has a large number of cascade observations and a few numbers of links. In this condition, DANI performs well. In all of the metrics, DANI has a better value than the other methods except in PWF and $\Delta\rho$, in which DANI results are comparable to DNE. Other methods also perform well on this dataset since there is a sufficient number of cascades for inferring a few strong links which are formed in this network. 

MemeTracker is our densest dataset with a smaller number of cascade observations. Since NetInf and MultiTree performance is sensitive to the observation size, they cannot perform well on this dataset. REFINE has poor performance on this dataset, and its results on metrics depending on detected communities by OSLOM are meaningless. DNE results are acceptable, but DANI performs better in terms of all metrics.

NBA has a large number of cascade observations and a relatively large number of links that have to be inferred. REFINE has poor performance on this dataset. The results of the other four methods are close together, but DNE has the best performance among them in link prediction, and in terms of metrics for structural properties, it is comparable with the best results. DANI is the only method in this experiment that does not produce isolated nodes. 

Twitter is our largest sparse dataset. DANI performs best on this dataset on all metrics except PWF and $\Delta$Cnd, in which DANI is running up method while DNE performs better. The poor performances of NetInf and MultiTree on Twitter indicate that they cannot perform well on sparse datasets.

Finally, we have the World dataset, which is similar to LinkedIn and NBA, except that it has a moderate number of observed cascades. DANI has the best performance in most metrics, While REFINE, unlike its low F1 in inference tasks, creates more similar average clustering coefficients.

\begin{table}[!ht]
\centering
\footnotesize
\caption{Result of DANI and other related works on real datasets}
\label{tab:realresult}
\begin{tabularx}{\textwidth}{Ycccccccccccc}
\hline\hline
                  & \textbf{Method} & \textbf{F1}   & \textbf{MF1}  & \textbf{mF1}  & \textbf{JS}   & \textbf{NMI}  & \textbf{PWF}  & \textbf{$\boldsymbol{\Delta}$ACS} & \textbf{$\boldsymbol{\Delta}$Cnd} & \textbf{$\boldsymbol{\Delta}$CN} & \textbf{$\boldsymbol{\Delta\rho}$} & \textbf{$\boldsymbol{\Delta}$CC}  \\ 
\hline\hline
\multirow{5}{*}{LinkedIn} & DANI            & \textbf{0.33} & \textbf{0.67} & \textbf{0.99} & \textbf{0.30} & \textbf{0.61} & 0.23          & \textbf{0.09}                     & \textbf{0.24}                     & \textbf{0.00}                    & 1.45                               & \textbf{0.62}                     \\ 
\cline{2-13}
                  & NetInf          & 0.30          & 0.65          & 0.98          & 0.30          & 0.60          & \textbf{0.30} & 0.26                              & 0.59                              & 0.05                             & \textbf{1.24}                      & 0.91                              \\ 
\cline{2-13}
                  & DNE             & 0.32          & 0.66          & 0.98          & 0.30          & 0.56          & 0.25          & 0.13                              & 0.52                              & 0.16                             & 1.49                               & 1.07                              \\ 
\cline{2-13}
                  & MultiTree       & 0.30          & 0.65          & 0.98          & 0.30          & 0.59          & 0.31          & 0.32                              & 0.62                              & 0.06                             & 1.26                               & 0.91                              \\ 
\cline{2-13}
                  & REFINE          & 0.29          & 0.63          & 0.98          & 0.70          & 0.38          & 0.14          & 2.76                              & 0.84                              & 0.64                             & 2.68                               & 2.19                              \\ 
\hline
\multirow{5}{*}{Meme} & DANI            & \textbf{0.99} & \textbf{0.99} & \textbf{0.99} & \textbf{0.40} & \textbf{0.68} & \textbf{0.93} & 0.16                              & \textbf{0.01}                     & \textbf{0.00}                    & \textbf{0.08}                      & \textbf{0.01}                     \\ 
\cline{2-13}
                  & NetInf          & 0.11          & 0.53          & 0.92          & 0.57          & 0.51          & 0.89          & 0.13                              & 0.37                              & 0.01                             & 0.68                               & 0.69                              \\ 
\cline{2-13}
                  & DNE             & 0.99          & 0.99          & 0.98          & 0.40          & 0.60          & 0.92          & 0.17                              & 0.02                              & 0.01                             & 0.15                               & 0.01                              \\ 
\cline{2-13}
                  & MultiTree       & 0.11          & 0.53          & 0.92          & 0.57          & 0.50          & 0.90          & \textbf{0.08}                     & 0.34                              & 0.01                             & 0.68                               & 0.69                              \\ 
\cline{2-13}
                  & REFINE          & 0.06          & 0.48          & 0.84          & 0.77          & 0.22          & 0.41          & 1.50                              & 0.05                              & 0.47                             & 0.24                               & 0.02                              \\ 
\hline
\multirow{5}{*}{NBA} & DANI            & 0.39          & 0.70          & 0.98          & 0.30          & \textbf{0.64} & 0.29          & 0.17                              & 0.23                              & \textbf{0.00}                    & 0.86                               & 0.34                              \\ 
\cline{2-13}
                  & NetInf          & 0.39          & 0.70          & 0.98          & 0.30          & 0.63          & 031           & \textbf{0.07}                     & 0.14                              & 0.03                             & 0.64                               & 0.60                              \\ 
\cline{2-13}
                  & DNE             & \textbf{0.41} & \textbf{0.70} & \textbf{0.98} & \textbf{0.29} & 0.63          & 0.30          & 0.18                              & \textbf{0.08}                              & 0.05                             & 0.81                               & 0.93                              \\ 
\cline{2-13}
                  & MultiTree       & 0.39          & 0.70          & 0.98          & 0.30          & 0.64          & \textbf{0.33} & 0.11                              & 0.10                    & 0.03                             & 0.66                               & 0.60                              \\ 
\cline{2-13}
                  & REFINE          & 0.01          & 0.49          & 0.96          & 0.70          & 0.23          & 0.08          & 0.85                              & 1.14                              & 0.65                             & \textbf{0.25}                      & \textbf{0.27}                     \\ 
\hline
\multirow{5}{*}{Twitter} & DANI            & \textbf{0.40} & \textbf{0.70} & \textbf{0.99} & \textbf{0.10} & \textbf{0.88} & 0.16          & \textbf{0.05}                     & 0.05                              & \textbf{0.00}                    & \textbf{1.12}                      & \textbf{0.46}                     \\ 
\cline{2-13}
                  & NetInf          & 0.01          & 0.50          & 0.99          & 0.75          & 0.11          & 0.01          & 5.71                              & 0.98                              & 0.92                             & 3.97                               & 0.78                              \\ 
\cline{2-13}
                  & DNE             & 0.35          & 0.67          & 0.99          & 0.10          & 0.85          & \textbf{0.21} & 0.13                              & \textbf{0.04}                     & 0.02                             & 1.30                               & 0.57                              \\ 
\cline{2-13}
                  & MultiTree       & 0.01          & 0.50          & 0.99          & 0.75          & 0.11          & 0.01          & 5.73                              & 0.99                              & 0.93                             & 3.98                               & 0.78                              \\ 
\cline{2-13}
                  & REFINE          & -             & -             & -             & -             & -             & -             & -                                 & -                                 & -                                & -                                  & -                                 \\ 
\hline
\multirow{5}{*}{World} & DANI            & \textbf{0.40} & \textbf{0.70} & \textbf{0.98} & 0.40          & \textbf{0.63} & 0.27          & 0.16                              & \textbf{0.05}                     & \textbf{0.00}                    & 0.97                               & 0.43                              \\ 
\cline{2-13}
                  & NetInf          & 0.30          & 0.65          & 0.97          & 0.37          & 0.61          & \textbf{0.28} & 0.05                              & 0.24                              & 0.03                             & 0.65                               & 0.57                              \\ 
\cline{2-13}
                  & DNE             & 0.39          & 0.69          & 0.98          & \textbf{0.35} & 0.61          & 0.28          & 0.17                              & 0.21                              & 0.05                             & 0.68                               & 0.81                              \\ 
\cline{2-13}
                  & MultiTree       & 0.30          & 0.65          & 0.97          & 0.37          & 0.61          & 0.31          & \textbf{0.03}                     & 0.26                              & 0.03                             & 0.63                               & 0.56                              \\ 
\cline{2-13}
                  & REFINE          & 0.01          & 0.49          & 0.49          & 0.70          & 0.25          & 0.13          & 1.00                              & 1.35                              & 0.66                             & \textbf{0.30}                      & \textbf{0.19}                     \\
\hline\hline
\end{tabularx}
\end{table}

\newpage
\section{Conclusion}\label{sec:conclusion}

Many studies have been conducted to infer networks using information diffusion, but they often neglect to preserve the properties of the underlying networks. Modularity is one of the most important features of social networks, and community detection algorithms require the network topology to extract the communities within the network. Therefore, these algorithms cannot be applied to networks derived from previously proposed network inference methods. Further, one of the main shortcomings of previous studies has been their time complexity.

In this paper, we proposed a novel method for inferring networks from diffusion data that maintains structural properties by using Markov transitions and node-node similarity from diffusion information. This claim was verified by comparing the topological structures of real and artificial networks using the proposed method. Moreover, we utilized three categories of metrics for evaluating the inference of links, assessing the similarity of inferred and original networks, and measuring structural characteristics for DANI and different related methods. 
The results showed that our algorithm accurately inferred networks while maintaining their topological structure. Furthermore, DANI requires less time to achieve the same or higher accuracy than competing methods. We showed that DANI performs better and faster in a more structured network and becomes slower when the modular structure fades to a random network. DANI can perform well on real social networks because they have modular topological structures. Furthermore, the performance of DANI is independent of the length of the cascade. Additionally, we designed a distributed version of the DANI algorithm in a MapReduce framework for real-world big data applications.

The inferred network can be far from the underlying network when missing data is ignored in diffusion observations \cite{Phantom2016}. Cascade sequences may contain missing data due to private accounts or incomplete crawls. Based on partial diffusion information, we will attempt to determine the hidden underlying network in a future study. 

\bibliographystyle{unsrtnat}
\bibliography{references}

\end{document}